\documentclass[10pt,amsmath,aps,amssymb,superscriptaddress,notitlepage,onecolumn,longbibliography]{revtex4-1}

\makeatletter
\renewcommand*{\p@subsubsection}{}
\makeatother

\usepackage{bm,amsmath,amssymb,
}
\usepackage{natbib} 
\usepackage{mydef}
\usepackage{amsmath}
\usepackage{graphicx}
\usepackage{mathtools}
\usepackage{mathtools}
\usepackage{float}
\usepackage{empheq}
\usepackage{stmaryrd}
\usepackage{epstopdf}
\usepackage{epstopdf}
\usepackage{amssymb}
\usepackage{wrapfig}
\usepackage{picture}
\usepackage{mwe,tikz}
\usepackage[linktoc=all]{hyperref}
\usepackage{color}
\usepackage{transparent}
\usepackage{textcomp}
\makeatletter
\newcommand{\verbatimfont}[1]{\def\verbatim@font{#1}}%
\makeatother

\definecolor{Lblu}{RGB}{193,193,234}
\definecolor{Mblu}{RGB}{131,131, 221}
\definecolor{Dblu}{RGB}{0,0, 255}
\definecolor{Lgr}{RGB}{192,192,192}
\definecolor{Mgr}{RGB}{138,138,138}

\definecolor{OR}{RGB}{244,127,31} 
\definecolor{BR}{RGB}{170,0,0} 
\definecolor{GR}{RGB}{26,211,26} 
\definecolor{VIO}{RGB}{170,0,170} 
\definecolor{PI}{RGB}{237,159,165} 

\def\XXint#1#2#3{{\setbox0=\hbox{$#1{#2#3}{\int}$}
		\vcenter{\hbox{$#2#3$}}\kern-.5\wd0}}

\begin{document}
\title{Cytoplasmic stirring by active carpets}

\author{Brato Chakrabarti}
\affiliation{Center for Computational Biology, Flatiron Institute, New York, NY 10010, USA}
\email{bchakrabarti@flatironinstitute.org}

\author{Stanislav Y. Shvartsman}
\affiliation{Center for Computational Biology, Flatiron Institute, New York, NY 10010, USA}
\affiliation{Department of Molecular Biology, Princeton University, Princeton, NJ 08544}
\affiliation{The  Lewis- Sigler  Institute  for  Integrative  Genomics,  Princeton  University, Princeton, NJ 08544}

\author{Michael J. Shelley}
\affiliation{Center for Computational Biology, Flatiron Institute, New York, NY 10010, USA}
\affiliation{Courant Institute, New York University, New York, NY 10012, USA}
\email{mshelley@flatironinstitute.org}

\date{\today}

\begin{abstract}
	Large cells often rely on cytoplasmic flows for intracellular transport, maintaining homeostasis, and positioning cellular components. Understanding the  mechanisms of these flows is essential for gaining insights into cell function, developmental processes, and evolutionary adaptability. Here, we focus on a class of self-organized cytoplasmic stirring mechanisms that result from fluid-structure interactions between cytoskeletal elements at the cell cortex. Drawing inspiration from streaming flows in late-stage fruit fly oocytes, we propose an analytically tractable \textit{active carpet} theory. This model deciphers the origins and three-dimensional spatio-temporal organization of such flows. Through a combination of simulations and weakly nonlinear theory, we establish the pathway of the streaming flow to its global attractor: a cell-spanning vortical twister. Our study reveals the inherent symmetries of this emergent flow, its low-dimensional structure, and illustrates how complex fluid-structure interaction aligns with classical solutions in Stokes flow. This framework can be easily adapted to elucidate a broad spectrum of self-organized, cortex-driven intracellular flows.
\end{abstract}
	\maketitle
	
	\section{Introduction}
	Cells in the biome generally fall within the 2-30 $\mu$m range \cite{milo2015cell}, primarily due to their dependence on inherently slow diffusion processes for intracellular transport. This suggests an evolutionary pressure to maintain the `right size' \cite{haldane1926being}. However, exceptions exist. From egg cells and algal plant cells to slime molds and unicellular ciliates,  cells can often reach several hundreds of microns or more in size. For these large cells, the transport and mixing of intracellular components by diffusion are extremely slow \cite{verchot2010cytoplasmic}. Instead, they often manifest persistent, large-scale intracellular flows  known as cytoplasmic streaming \cite{van2008nature, yi2011dynamic, almonacid2015active, hird1993cortical, kimura2017endoplasmic}.
		
	The stirring of cytoplasm is most commonly orchestrated by  cytoskeletal elements at the cell cortex. This class of boundary driven flows  can be  broadly divided into two categories. First, in which the active processes that generate these flows are predetermined during developmental stages. This results in a unidirectional coupling between cortical stresses and cytoplasmic flows. Notable examples of this process include the actomyosin driven reorganizations seen in \textit{Xenopus} egg extract \cite{sakamoto2020tug,ierushalmi2020centering,shamipour2021cytoplasm},  cytoplasmic streaming observed in algal cells \cite{corti1774osservazioni,woodhouse2013cytoplasmic,verchot2010cytoplasmic}, and the intracellular flows in amoeba spurred by cellular deformations \cite{koslover2017cytoplasmic}. Here, we focus on the second category of cytoplasmic stirring that arises through the dynamical self-organization of the cytoskeletal elements driven by bidirectional feedback between bulk flows and cortical stresses. Two prominent examples of this are streaming flows in the oocytes of \textit{C. elegans} \cite{kimura2017endoplasmic} and the fruit fly, \textit{Drosophila melanogaster} \cite{gutzeit1982time, glotzer1997cytoplasmic, quinlan2016cytoplasmic, ganguly2012cytoplasmic, khuc2015cortical}. In both of these examples, the flow is believed to emerge from fluid-structure interactions where nanometric molecular machines like Kinesin-1 traverse along flexible microtubules (MTs) anchored on the cell cortex, carrying payloads that entrain fluid \cite{lu2023go, lu2016microtubule, monteith2016mechanism}. In the fruit fly, the interplay of entrained flow and collective deformation of anchored MT beds drives a self-organized vortical flow that spans the entire cell chamber of $100-300$ \ $\mu$m \cite{monteith2016mechanism}.
	
	To understand this self-organized flow in fruit flies, one study developed an active Brinkman-elastica model that describes the dynamics of an immersed and flexible microtubule bed upon which agents carry payloads \cite{stein2021swirling}. Special homogeneous solutions to this model, set in a 2D disk, showed that a global bending instability led to a large-scale vortex within the cavity. Recent large-scale computations of discrete motor-loaded microtubules set in 3D geometries further elucidated the flow topologies \cite{dutta2023self}. In the appropriate parameter regime, these simulations involving over a million degrees of freedom, consistently produce cell-spanning vortical flows, known as twisters, with remarkable robustness. Mathematically, this dynamics suggests that intracellular twisters are stable global attractors of a high-dimensional dynamical system. That said, how this complex fluid-structure interaction phenomenon converges to such a robust solution remains unknown. Here, we address this question by developing an \textit{active carpet} model that is analytically tractable and is suitable for embedding in 3D cellular geometries. Drawing inspiration from the intricate flows observed in the fruit fly, this model possesses the versatility to be tailored for elucidating a diverse range of cortically driven, self-organized cytoplasmic flows.

	\begin{figure*}
		\centering
		\includegraphics[width=1\textwidth]{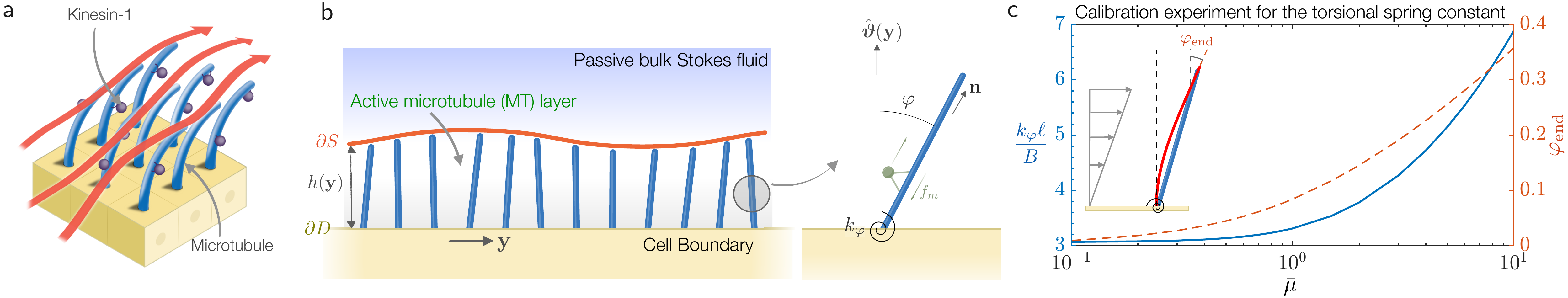}
		\caption{\textbf{\textsf{Schematic of the experimental biological system, coarse-grained theory, and relevant approximations.}} $\boldsymbol{\mathsf{a}}$, A magnified view on the cell boundary depicts the bent microtubules (MTs) in blue and molecular motors (kinesin-1) in purple, and the streaming flow in red. $\boldsymbol{\mathsf{b}}$, A schematic illustrates the coarse-grained active carpet model of the streaming flow. MTs are approximated as rigid rods with a torsional spring at their anchoring point to the cell boundary. $\boldsymbol{\mathsf{c}}$, We calibrate the spring stiffness $k_\vphi$ using the deflection of a clamped MT fiber in a simple shear flow, approximated as a rigid rod of the same length $L$. The flow-strength is characterized by $\bar{\mu} = 8 \pi \mu \dot{\gamma} L^4/(B c)$, where $\mu$ is the fluid viscosity, $\dot{\gamma}$ is the shear rate, $B$ is MT's bending rigidity, and $c = -\log(\varepsilon^2 \mathrm{e})$ is a geometric constant associated with the fiber aspect ratio $\varepsilon$. As  $\bar{\mu}$ increases, so does the spring stiffness. We use the mean value over a range of $\bar{\mu}$ for which the filament deflections are moderate, resulting in $k_\vphi \approx 4-5 B/L$. }
		\label{fig:Fig1}
	\end{figure*}

\section{Active carpet model}
	
	\noindent Our model approximates semiflexible MT filaments as rigid rods of length $L$  anchored with their minus-ends to the cell cortex.  At their anchoring point, these rods have a torsional spring that mimics the bending response of clamped, flexible filaments  \cite{kimura2017endoplasmic,pellicciotta2020cilia,stein2021swirling}. We systematically perform this coarse-graining (see Fig.~\ref{fig:Fig1}) by callibrating the bending response of a flexible filament to rotations of an anchored rigid rod in a simple shear flow. This allows us to quantitatively map the torsional spring constant to the properties of the MT filament.  Each rod acts as a track for the Kinesin-1 motor proteins, as illustrated in  Fig.~\ref{fig:Fig1}$\boldsymbol{\mathsf{b}}$. The forces from the plus-end directed cargos, propelled by these motor proteins, are coarse-grained into a uniform compressive force density, $f_m$. This force-density is directed along the bound MT towards its anchored minus end. Following Newton's third law, the motors exert an equal and opposite force on the cytoplasmic fluid. We characterize this dense MT carpet  \cite{quinlan2016cytoplasmic} by a uniform areal density $c_0$ and its orientation by a polarity field $\bn(\by,t)$, where $\by \in \partial D$ is the surface coordinate  on the cell boundary  (see Fig.~\ref{fig:Fig1}). In this paper, we will denote any surface point by $\by$ and any point within the cytoplasmic volume by $\bx$. The polarity field evolves according to Jeffery's equation  \cite{leal2007advanced} as 
	\begin{equation}\label{eq:polarity}
		\partial_t \bn(\by,t)  = (\bI - \bn \bn) \cdot \nabla \bu \Big|_{\partial D} \cdot \bn + \frac{k_\vphi}{\xi_r} \bT_0 \times \bn.
	\end{equation}
	Here $\bu(\bx,t)$ is the emergent cytoplasmic fluid velocity that needs to be determined. The second term in Eq.~\eqref{eq:polarity}  characterizes the response from the restoring torque on the rod. Here, $\xi_r = 4 \pi \mu L^3/3 \log(2/\varepsilon)$ with  $\varepsilon \approx 10^{-3}$ being the apsect ratio of the rod and $\mu \sim 1$ Pa.s being the viscosity of the cytoplasm, which we model as an incompressible Newtonian fluid \cite{ganguly2012cytoplasmic}.  The restoring torque $k_\vphi \bT_0$ stems from the torsional spring of stiffness $k_\vphi$ and penalizes deviation of the MT orientation from the inward unit surface  normal $\hat{\boldsymbol{\vartheta}}(\by)$. We estimate $k_\vphi \sim 4-5 B/L$, where $B \approx 20$ pN.$\mu$m$^2$ \cite{howard2002mechanics} is the stiffness of a single MT fiber (see Fig.~\ref{fig:Fig1}$\boldsymbol{\mathsf{c}}$).  The action of the molecular motors that drive the internal  flow is approximated by a concentrated layer of stress jump $\bar{\bff}(\by,t)$ across the interface $\partial S$ delineated by the tips of the MT layer (see Fig.~\ref{fig:Fig1}$\boldsymbol{\mathsf{b}}$).  Here we disregard the surface fluctuations of $\partial S$ and define it as $\partial S := \partial  D + L \hat{\boldsymbol{\vartheta}}(\by)$. The quantitative impact of this simplification on the emergent dynamics is verified a posteriori.  In the Stokesian limit of interest, the dimensionless governing equation for the momentum balance is then given as
	\begin{alignat}{2}
		-\nabla q + \Delta \bu = \mathbf{0}, &  \ \ \nabla \cdot \bu  = 0, \label{eq:homst}& \\ 
		\bu(\by,t) = \mathbf{0}, &  \ \ \text{ on } \partial D, & \\ 
		\llbracket \bu \rrbracket = \mathbf{0}, \ \ \llbracket \boldsymbol{\sigma}  \cdot \hat{\boldsymbol{\vartheta}} \rrbracket (\by,t) = -\bar{\bff}(\by,t).& \ \ \text{ on } \partial S. & \label{eq:stokeappx}
	\end{alignat}
	Here $q$ is the fluid pressure, $\llbracket a \rrbracket = a|_{\partial S^+} - a|_{\partial S^-}$ denotes the jump of any variable across the interface $\partial S$, and $\boldsymbol{\sigma} = -q \bI + (\nabla \bu + \nabla \bu^T)$ is the Newtonian stress tensor.

	Using the MT length $L$ as the characteristic length scale, the relaxation time $\tau_r \sim \xi_r/k_\vphi$ as the time scale and a  viscous force scale, the parameters in the problem can be combined to yield two key dimensionless control parameters: (i) $\bar{\rho} =  c_0 L^2$ characterizing the surface density of the MT bed and (ii) $\bar{\sigma} = \xi_r  f_m/(\mu L k_\vphi)$ serving as a measure of activity. This dimensionless activity compares  the relative strength of compressive motor forces to those derived from the restoring spring. There are two additional geometric parameters in the problem: (i) $\mathcal{R} = L_s/L > 1$ is the ratio of the system size to the MT length and (ii) $\chi = 2 \pi/\log(2/\varepsilon)$ is a geometric constant that depends weakly on the aspect ratio $\varepsilon$ of the MT. With this scaling, the dimensionless traction jump in Eq.~\eqref{eq:stokeappx} can be expressed as (see SI)
	\begin{equation}\label{eq:jump}
		\bar{\bff}(\by,t) =  \bar{\rho} \bar{\sigma} \mathbf{n} + \bar{\rho} \chi \left[ \mathbf{T}_0 \times \mathbf{n}-\frac{\mathbf{n n n}}{2}: \nabla \mathbf{u}\Big|_{\partial D} \right]. 
	\end{equation}
	The first term in Eq.~\eqref{eq:jump} arises from the forces exerted  by the Kinesin-1 motor proteins on the fluid (see Fig.~\ref{fig:Fig1}$\boldsymbol{\mathsf{b}}$). The second term represents forces stemming both from the torsional spring and from the rod's response in the mean-field fluid flow (see SI). 	 This completes our active carpet theory. In essence, our model integrates a boundary force field as defined by Eq.~\eqref{eq:jump} to an internal homogeneous Stokes flow delineated in Eqs.~\eqref{eq:homst}-\eqref{eq:stokeappx}. The forcing and the emergent flow evolve self-consistently with a partial differential equation (PDE) that details the polarity of the MT bed as laid out in Eq.~\eqref{eq:polarity}. We highlight that in contrast with other problems of self-organized flows \cite{keber2014topology,bell2022active}, the dynamics of the polarity field does not involve any surface transport on the cell-cortex. Its evolution is only coupled to the  cytoplasmic flows through the velocity gradient on the cell boundary as seen in Eq.~\eqref{eq:polarity}.

	\section{Results}
	
	\subsection{Streaming flows in half-space}
\begin{figure*}
		\centering
		\includegraphics[width=1\textwidth]{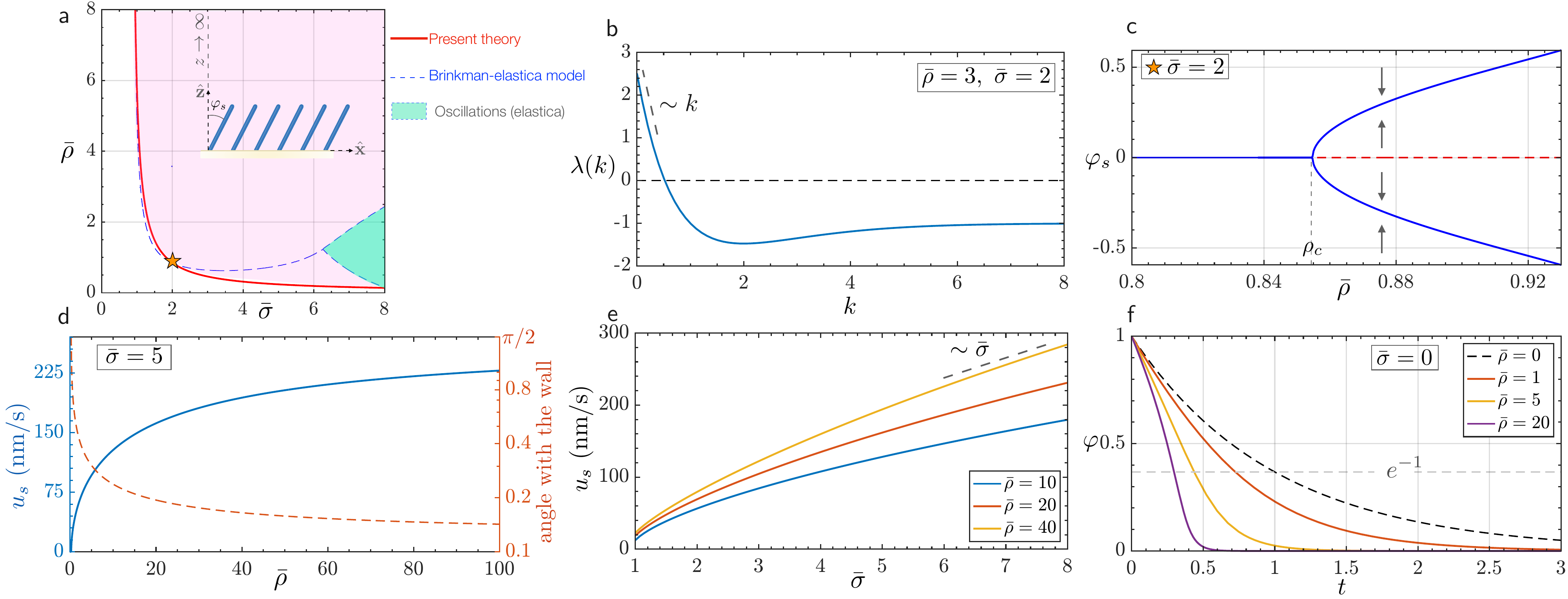}
		\caption{\textbf{\textsf{Bifurcation analysis, streaming speeds, and relaxation dynamics in half-space.}} $\boldsymbol{\mathsf{a}}$, Linear stability analysis (LSA) of the continuum theory predicts the emergence of a streaming state characterized by a homogeneous deflection of the MT bed (shown schematically). Overlaid on the stability boundary (depicted in red) are results from an active Brinkman-elastica model (in blue) obtained from \cite{stein2021swirling}. Our active carpet model does not capture the oscillatory regime of the elastic filaments but agrees well for the prediction of streaming states. $\boldsymbol{\mathsf{b}}$, Dependence of the growth rate $\lambda$ on the plane wave number $k$. The growth rate is maximal for $k=0$ indicating a long-wavelength instability and decreases linearly with $k$ for small wavenumbers. Large-wavenumbers are exponentially damped and asymptotes to $\lambda = -1$. $\boldsymbol{\mathsf{c}}$, Steady-state deflection of the fiber bed as a function of $\bar{\rho}$. As evidenced by the bifurcation diagram, the streaming states arise from a supercritical pitchfork bifurcation of the straight MT bed.  $\boldsymbol{\mathsf{d}}$, Dimensional streaming speed $u_s$ and the deflection of the MT bed relative to the wall as a function of $\bar{\rho}$. As $\bar{\rho} \to \infty$ the streaming speed asymptotes to a finite value. $\boldsymbol{\mathsf{e}}$, At a given density $\bar{\rho}$, the streaming speed increases monotonically with activity. For large density we find $u_s \sim \bar{\sigma}$. $\boldsymbol{\mathsf{f}}$, Relaxation dynamics of a MT bed for various $\bar{\rho}$. The characaterisitc relaxation time of an isolated MT is $\tau_r \sim \xi_r/k_\vphi$. In the presence of hydrodynamic interactions, the bed relaxes faster, indicating the emergence of a collective relaxation time $\tau_c \sim \tau_r/\bar{\rho}$.}
		\label{fig:Fig2}
	\end{figure*}

	We initiate our exploration with simple, yet insightful dynamics within a 2D half-space. In this setup, an infinite planar array of fibers is anchored to a no-slip wall located at $z=0$ (see Fig.~\ref{fig:Fig2}$\boldsymbol{\mathsf{a}}$). A fixed point for the system occurs when all the fibers are aligned along the $z$ direction. In this configuration,  represented as $\bn_0 = \uvc{z}$, the motor-induced pushing forces renormalize the mean-field pressure, yielding $\bu_0 = \mathbf{0}$. A small homogeneous transverse perturbation in the polarity field of the form $\bn = \uvc{z} + \epsilon \vphi \uvc{x}$ ($\epsilon \ll 1$) evolves as,
	\begin{equation}\label{eq:lshs}
		\partial_t \vphi = \left[\bar{\rho} \bar{\sigma}-\left(\bar{\rho} \chi+1\right)\right] \varphi.
	\end{equation}
	Here $\vphi$ is the angle made by the rods with the $\uvc{z}$ axis. The first term in Eq.~\eqref{eq:lshs} is destabilizing and arise from an unidirectional shear flow in the active MT layer that tries to rotate the rods away from their upright position. The stabilizing second term is due to the restoring torque from the torsional spring. As evident from Eq.~\eqref{eq:lshs}, the fixed point  becomes unstable when,
	\begin{equation}\label{eq:critac}
		\bar{\sigma} > \left[\frac{1}{\bar{\rho}} + \chi\right].
	\end{equation}
	This stability boundary, beyond which we observe steady streaming states, is shown in red on Fig.~\ref{fig:Fig2}$\boldsymbol{\mathsf{a}}$. A plane-wave perturbation of the form $\epsilon \vphi(x) \sim \epsilon \hat{\vphi}_k e^{\mi k x + \lambda t}$, elucidates the wavenumber dependence of the growth rate as
	\begin{equation}\label{eq:disp}
		\lambda(k) = -1+\bar{\rho} e^{-k}(1-k)(\bar{\sigma}-\chi).
	\end{equation} 
	The above dispersion relation accentuates the long-wavelength nature of the instability. We find that the homogeneous state, characterized by $k=0$ is the most unstable wavenumber. This mirrors behaviors observed in instabilities of active suspensions \cite{saintillan2013active}. Further, from Eq.~\eqref{eq:disp}, we observe that large wavenumbers experience exponential damping, and the growth rate asymptotes to $\lambda(k\to \infty) \to -1$ (see Fig.~\ref{fig:Fig2}$\boldsymbol{\mathsf{b}}$). This is in stark contrast to active-gel theories \cite{julicher2018hydrodynamic}, where nematic elasticity results in an algebraic decay of the growth rate at high-wavenumbers. 
	
	Nonlinear simulations further confirm that the emergent streaming state is homogeneous, characterized by a constant deflection angle $\vphi_s$ of the MT bed with the $\uvc{z}$ axis. Figure~\ref{fig:Fig2}$\boldsymbol{\mathsf{c}}$ illustrates this steady-state deflection angle for a given activity $\bar{\sigma}$, plotted against the MT bed density. We find, $\vphi_s \sim \sqrt{\bar{\rho} - \rho_c}$, close to the critical density $\rho_c$, indicating the hallmarks of a supercritical pitchfork bifurcation. In this emergent steady-state, the fluid flow is constant and unidirectional above the MT layer and is characterized by the streaming speed $u_s$. At a given activity, the streaming speed  increases with the bed density and is accompanied by increasing alignment of the MT fibers with the wall (see Fig.~\ref{fig:Fig2}$\boldsymbol{\mathsf{d}}$). Nonetheless, at large $\bar{\rho}$, hydrodynamic interactions between MTs become screened suggesting that the energy injected by the motor-proteins are dissipated at the scale of a single rod, resulting in a scaling of the form $\mu u_s^2 \sim f_m u_s$. This scaling law leads to a streaming speed independent of the MT density and increasing linearly with the motor force alone. Indeed, these observations are corroborated by the saturation of $u_s$ with increasing density on Fig.~\ref{fig:Fig2}$\boldsymbol{\mathsf{d}}$ and its variation with the activity on Fig.~\ref{fig:Fig2}$\boldsymbol{\mathsf{e}}$. Finally, our theory underscores the emergence of a fast relaxation time of the MT bed, arising from the collective hydrodynamic interactions. Analyzing the relaxational dynamics of an initially deflected bed (see Fig.~\ref{fig:Fig2}$\boldsymbol{\mathsf{e}}$) we determine this collective response time scale as $\tau_c \sim \tau_r/\bar{\rho}$, with $\tau_r \sim \xi_r/k_\vphi$ representing the single MT relaxation time. With these insights, we now ask, how this instability and the emergent flows are organized in an abstracted spherical egg cell, where topology prevents homogeneous solutions.
	
	\begin{figure*}
		\centering
		\includegraphics[width=1\textwidth]{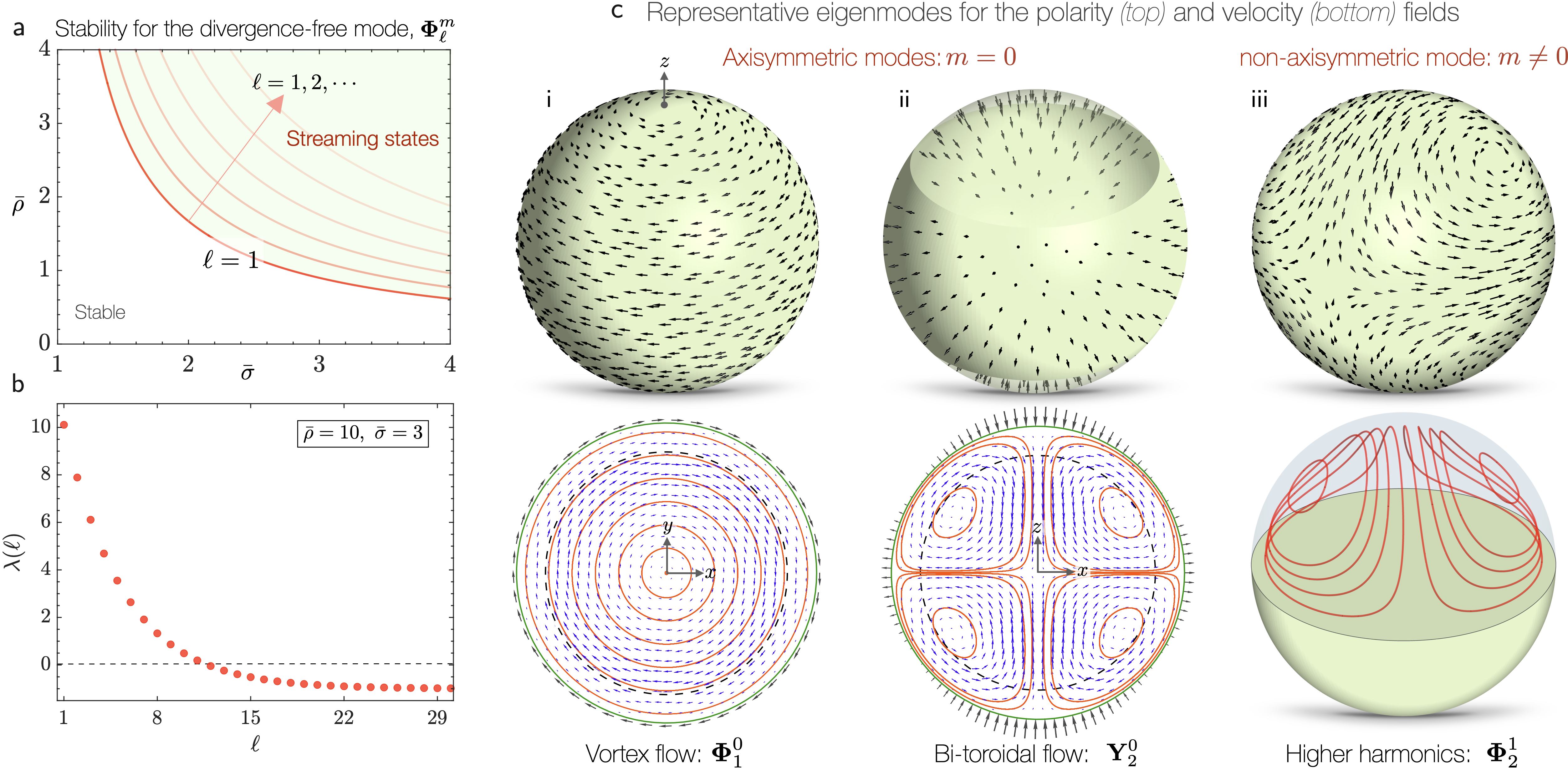}
		\caption{\textbf{\textsf{Linear stability analysis of streaming flows inside a spherical egg.}} $\boldsymbol{\mathsf{a}}$, Stability boundary illustrating the onset of twisters within a sphere. The threshold for the instability is delineated by $\ell = 1$,  representing the most unstable wavenumber. Successive curves in progressively lighter shades of red denote increasing discrete wavenumbers as they become unstable. $\boldsymbol{\mathsf{b}}$, Variation of the growth rate $\lambda(\ell)$ with the polar wavenumber $\ell$ for a given MT density $\bar{\rho} = 10$ and activity $\bar{\sigma} = 3$. This  highlights the long-wavelength nature of the instability. High wavenumbers are exponentially damped and asymptotes to $-1$ as $\ell \to \infty$.  $\boldsymbol{\mathsf{c}}$, Top row: Three representative surface polarity fields in terms of the VSH basis. The dominant unstable eigenmodes are characterized by $\boldsymbol{\Phi}_\ell^m$ and the subdominant one is given by $\boldsymbol{\Psi}_\ell^m$ (not shown). Bottom row: Corresponding interior flows with the dashed line in circular cross-section representing the surface $\partial S$. Axisymmetric modes with $m=0$ exhibit a cell-spanning vortex (i) and a bitoroidal flow (ii). The non-axisymmetric mode in (iii) illustrates saddles and a center in the polarity field. The associated flow-field is depicted for the upper hemisphere due to its mirror symmetry in the lower hemisphere. Here, we have fixed $R = 5$ corresponding to a dimensional radius of $\sim 100$ \ $\mu$m.}
		\label{fig:Fig3}
	\end{figure*}
	
	\subsection{Emergent vortex in a sphere: linear stability analysis}
	 To build towards our theoretical and computational investigation of 3D interior flows, we first highlight that solutions to Eqs.~\eqref{eq:homst}-\eqref{eq:stokeappx} can be written as \cite{pozrikidis1992boundary,tanzosh1992boundary}
	\begin{equation}\label{eq:intfor}
		\begin{split}
			\bu(\bx) = &\frac{1}{8 \pi} \int_{\partial S} \bar{\bff}(\by) \cdot \mathcal{G}(\bx,\by) \ \md A(\by)  - \frac{1}{8 \pi} \int_{\partial D} \bff^w(\by) \cdot \mathcal{G}(\bx,\by) \ \md A(\by),
		\end{split}
	\end{equation}
	where $\mathcal{G}(\bx,\by)$ is the free-space Green's function for the Stokes equation. In Eq.~\eqref{eq:intfor},
	$\bar{\bff}(\by)$ is the traction jump across $\partial S$ specified by Eq.~\eqref{eq:jump} and $\bff^w(\by)$  is an unknown traction on the cell cortex $\partial D$. It is determined by imposing the no-slip condition $\bu(\by) = \mathbf{0}$ on $\partial D$. This boundary-integral formulation allows fast simulations of flows inside smooth closed surfaces and helps us to derive analytical solutions for spherical egg-cells as demonstrated below. 
	
	We start  by examining the emergence of streaming flows within a sphere of dimensionless radius 
	$R > 1$. Analogous to the half-space analysis, a fixed-point of the system with  $\bu = \mathbf{0}$  arises when $\bn_0 = -\uvc{e}_r$, where $\uvc{e}_r$ is the unit-outward normal to the sphere. To probe the stability of this state, we consider the evolution of a slightly perturbed polarity field $\bn = -\uvc{e}_r + \epsilon \tilde{\bn}(\theta,\phi) e^{\lambda t} \ (\epsilon \ll 1)$, where $\tilde{\bn}$ lies in the tangent-space of the sphere. Here, $(r,\theta,\phi)$ represents the spherical coordinates, where $\theta \in [0,\pi]$ is the polar and $\phi \in [0, 2 \pi)$ is the azimuthal angle. Realizing that the associated flow $\bu \sim \mathcal{O}(\epsilon)$, the evolution of the perturbation from the dimensionless form of the linearized Eq.~\eqref{eq:polarity} is given as
	\begin{equation}\label{eq:linsp}
		\lambda \tilde{\bn} = -\partial_r \bu \big|_{r=R} - \tilde{\bn}.
	\end{equation}
	As before, the first term in Eq.~\eqref{eq:linsp} is associated with a de-stabilizing shear-flow in the MT layer and the second term arises from the restoring torque. Drawing  parallel to our plane-wave expansion in the half-space, we expand $\tilde{\bn}$ on the basis of vector spherical harmonics (VSH) as
	\begin{equation}\label{eq:nvsh}
		\tilde{\bn}(\theta,\phi) = \sum_{\ell \geq 1, |m| \leq \ell} b_{\ell}^m  \boldsymbol{\Psi}_\ell^m  + c_{\ell}^m \boldsymbol{\Phi}_\ell^m.
	\end{equation}
	Here we have introduced the set $\{\mathbf{Y}_\ell^m, \boldsymbol{\Psi}_\ell^m, \boldsymbol{\Phi}_\ell^m \}$, which are three orthogonal VSH of degree $\ell$ and order $m$ (for $|m| \leq \ell$) with $\ell$ being the polar and $m$ the azimuthal wavenumber \cite{hill1954theory} (see Appendix~\ref{app:VSH}). Given a vector field represented in this basis, $\mathbf{Y}_\ell^m$ delineates its radial variation, $\boldsymbol{\Psi}_\ell^m$ accounts for the curl-free  and $ \boldsymbol{\Phi}_\ell^m$  the divergence free component.
	
	The Ansatz in Eq.~\eqref{eq:nvsh} ensures that $\tilde{\bn}$ is on the tangent space and the coefficients $\{ b_{\ell}^m, c_{\ell}^m\}$ correspond to their specific amplitudes. We now recall that the integral operators defined by Eq.~\eqref{eq:intfor} diagonalize in the VSH basis \cite{schmitz1982creeping,corona2018boundary,kim2013microhydrodynamics}. This observation motivates a solution approach in which we first expand the tractions in Eq.~\eqref{eq:intfor} on the chosen basis. On using the no-slip boundary condition we can then determine the unknown velocity $\bu(\bx)$ in terms of $\tilde{\bn}$. This allows us to obtain an analytical expression (see SI) for the growth rate for each polar wavenumber $\ell$ as
	\begin{equation}\label{eq:lambda}
		\lambda(\ell) = -1 +  \left(\frac{R-1}{R}\right)^{\ell+2} \bar{\rho} (\bar{\sigma}-\chi).
	\end{equation}
	
	This expression is the central result from our linear theory and elucidates various facets of the instability. It predicts that the homogeneous fixed point becomes unstable for increasing activity or MT density and exhibits a weak dependency on the sphere size (see Fig.~\ref{fig:Fig3}$\boldsymbol{\mathsf{a}}$). Notably, as $R \to \infty$, we recover the stability threshold of the half-space, described by Eq.~\eqref{eq:critac}. We further highlight that the dispersion relation is only a function of the quantized polar wavenumber $\ell$ and is independent of the azimuthal wavenumber $m$. This follows from the spectral properties of the integral operators as discussed in the SI. 
	
	It is further evident from Eq.~\eqref{eq:critac} that $\ell = 1$ emerges as the fastest growing mode and defines the stability boundary as seen in  Fig.~\ref{fig:Fig3}$\boldsymbol{\mathsf{a}}$. This mirrors the long-wavelength dynamics observed in the half-space as large polar wavenumbers are exponentially damped with $\lambda(\ell)\to -1$ as $\ell \to \infty$ (see Fig.~\ref{fig:Fig3}$\boldsymbol{\mathsf{b}}$).  The dominant unstable eigenmodes are characterized by the divergence-free field $\boldsymbol{\Phi}_\ell^m$. Figure~\ref{fig:Fig3}$\boldsymbol{\mathsf{c}}$(i) shows the  axisymmetric polarity field for the most unstable wavenumber $\ell = 1$. The associated fluid flow is reminiscent of a system-size \textit{vortex}. As seen from Fig.~\ref{fig:Fig3}$\boldsymbol{\mathsf{c}}$(i), this flow is characterized by a pure rigid-body rotation in the interior fluid and a shear flow in the MT layer, conforming to the no-slip boundary condition on the cell-cortex. 
	
	\subsection{Nonlinear dynamics and the \textit{twister} flow}

	Building on our understanding from the linear theory, we now probe the full nonlinear dynamics to explore the self-organization of emergent flows beyond the instability threshold. To this end, we start from a slightly perturbed base state and numerically evolve Eqs.~\eqref{eq:polarity}, \eqref{eq:jump}, and \eqref{eq:intfor} simultaneously (see Appendix~\ref{app:NM}).  Figure~\ref{fig:Fig4}$\boldsymbol{\mathsf{a}}$ shows the evolution of the polarity field over time (portrayed on the exterior of the sphere). We characterize this evolution by color coding the surface by the local polar order-parameter defined as 	
	\begin{equation}\label{eq:OP}
		P(\theta,\phi) = \left\Vert\left(\bI - \uvc{e}_r \uvc{e}_r\right) \cdot \bn(\theta,\phi)\right\Vert.
	\end{equation}
	As evident from Eq.~\eqref{eq:OP}, regions of low polar order correspond to MTs being orthogonal to the cell surface. To further quantify the polarity field's spatio-temporal features, we seek a spectral expansion of the form $\bn(\theta,\phi) = a_{\ell}^m  \mathbf{Y}_\ell^m + b_{\ell}^m  \boldsymbol{\Psi}_\ell^m  + c_{\ell}^m \boldsymbol{\Phi}_\ell^m$. This approach allows us to  introduce the power-spectrum, $\mathcal{P}_\ell(t)$ (see Appendix~\ref{app:VSH}), which quantifies the azimuthally averaged energy content  across all modes of a given polar wavenumber $\ell$.

		\begin{figure*}
		\centering
		\includegraphics[width=1\textwidth]{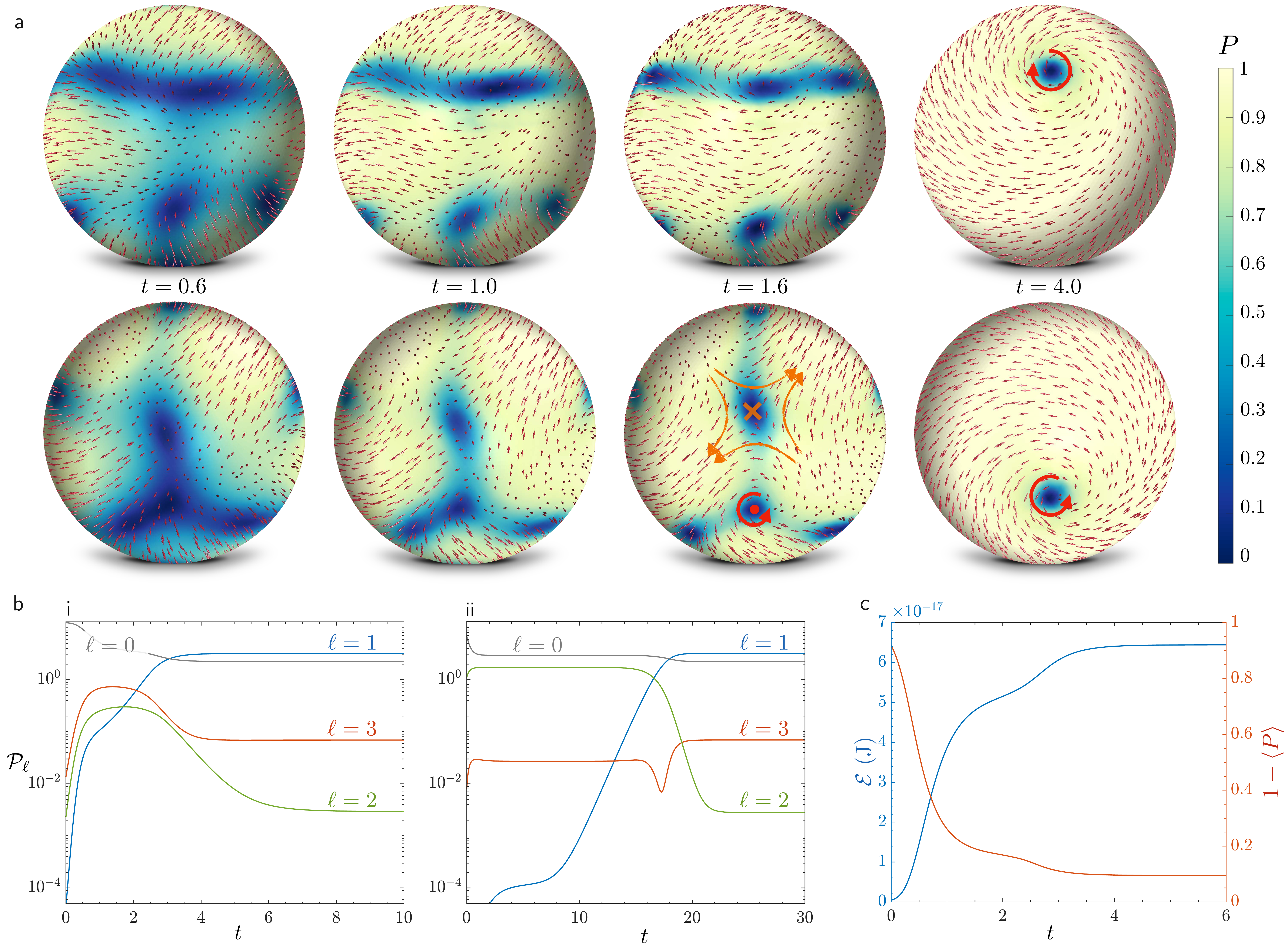}
		\caption{\textbf{\textsf{Nonlinear dynamics and the pathway to the robust swirling state.}} $\boldsymbol{\mathsf{a}}$, Multiple views of the snapshots from the evolution of the polarity field $\bn$ are displayed by arrows on the exterior surface of the sphere. The surface is color coded by the local polar order parameter $P(\theta,\phi)$. Emergent topological defects with charge $-1$ (saddle) are marked by $(\boldsymbol{\times})$ and those with charge $+1$ (center) is marked by $\boldsymbol{\circ}$. Time-stamps are in terms of the dimensionless time scaled by the single MT relaxation time $\tau_r$. $\boldsymbol{\mathsf{b}}$, Evolution of the power spectrum of the polarity field under two different initial conditions. The base state of $\bn_0 = -\uvc{e}_r$ corresponding to $\ell = 0$ was perturbed by (i) including contributions from $\ell = 1,2,3$ and (ii) from only $\ell = 2$ mode in the initial data. The mode $\ell =1$ is associated with the emergent streaming flows. $\boldsymbol{\mathsf{c}}$, The evolution of the dimensional elastic energy $\mathcal{E}$ alongside the average-polar order parameter $\langle P \rangle$. The increase in the bending energy is accompanied by a decrease in $1 - \langle P\rangle$ characterizing the aligned state of MTs. Parameters: $\bar{\rho} = 10$ (corresponding to roughly $3 \times 10^3$ MTs on the surface), $\bar{\sigma} = 3, \chi = 0.83$, and $R = 5$. The choice of $\bar{\sigma}$ corresponds to a dimensional motor force denisty of $f_m \sim 0.07$ pN/$\mu$m. Assuming a single motor force to be $\mathcal{O}(2$ pN) \cite{visscher1999single} leads to an estimate of 1-3 kinesin motors per MT \cite{stein2021swirling}.}
		\label{fig:Fig4}
	\end{figure*}
	In Fig.~\ref{fig:Fig4}$\boldsymbol{\mathsf{a}}$, we observe that starting from the nearly radial initial condition, the polarity field rapidly develops regions of high polar order over the collective response timescale, $\tau_c$.  These regions are indicative of well-aligned MT patches inducing flows parallel to their local orientations.  This transient dynamics is confirmed by a marked reduction in the radial component associated with $\mathcal{P}_0$, accompanied by increased contributions from other wavenumbers in the power-spectrum, as shown in Fig.~\ref{fig:Fig4}$\boldsymbol{\mathsf{b}}$(i).  As the cytoplasmic flow ensues, it drives a self-organization across the system in which regions of high polar order compete and interact. These interactions progressively lead to the formation of pronounced low polar order patches and give birth to topological defects of charges both $+1$ (center) and $-1$ (saddles) as depicted in Fig.~\ref{fig:Fig4}$\boldsymbol{\mathsf{a}}$. Always appearing in pairs, these disclination points in the surface polarity field preserve a global topological charge of two, satisfying the Poincar\'e-Hopf theorem. Ultimately, the system settles into a steady state, in which regions of high polar order coalesce, leaving us with an axisymmetric swirling state which we call a \textit{twister} following \cite{dutta2023self}. As evident in the final snapshot, the alignment axis connects the two $+1$ defect centers on the two poles. The orientation and the chirality of the emergent field is determined by initial conditions. Figure~\ref{fig:Fig4}$\boldsymbol{\mathsf{b}}$(i) reveals that this final state is characterized by a dominant global contribution from $\ell = 1$ mode corresponding to a \textit{vortex} flow in the interior and weaker contributions from $\ell = 2, 3$. We further highlight in Fig.~\ref{fig:Fig4}$\boldsymbol{\mathsf{c}}$, that the evolution towards this twister state is paired with a monotonic increase in the bending energy of the MT bed defined as
	\begin{equation}
		\mathcal{E} = \frac{1}{2} k_\vphi \int_{\partial D} \left[\cos^{-1}\left(-\uvc{e}_r \cdot \bn(\theta,\phi)\right)\right]^2 \ \md A.
	\end{equation}
	This increase  stems from the collective bending of the MTs induced by the emergent cytoplasmic flows. In the steady-state, the energy injected by the molecular motors is balanced by the viscous dissipation in the cytoplasmic flows, which in turn maintain the bent conformation of the MT layer. 
	
	We now ask, how robust is this axisymmetric steady-state? We test this by initiating the system with a different initial condition. The evolution of $\mathcal{P}_\ell(t)$ in Fig.~\ref{fig:Fig4}$\boldsymbol{\mathsf{b}}$(ii) reveals that for this initial condition, the system quickly settles onto a low-dimensional manifold with a dominant contribution from the $\ell = 2$ mode. Our simulations indicate that this quasi-steady state closely aligns with the non-axisymmetric eigenmode shown in Fig.~\ref{fig:Fig3}$\boldsymbol{\mathsf{c}}$(iii). It stands distinct from the transient observed in the previous case, as seen in Fig.~\ref{fig:Fig4}$\boldsymbol{\mathsf{b}}$(i). However, over long time, the mode associated with the vortex flow ($\ell = 1$) grow, accompanied by the decrease of the $\ell = 2$ mode as before. The culmination is an analogous axisymmetric twister state, characterized by an identical power-spectrum, which underscores the possibility of a single global attractor governing the collective dynamics.

	\begin{figure*}
		\centering
		\includegraphics[width=0.94\textwidth]{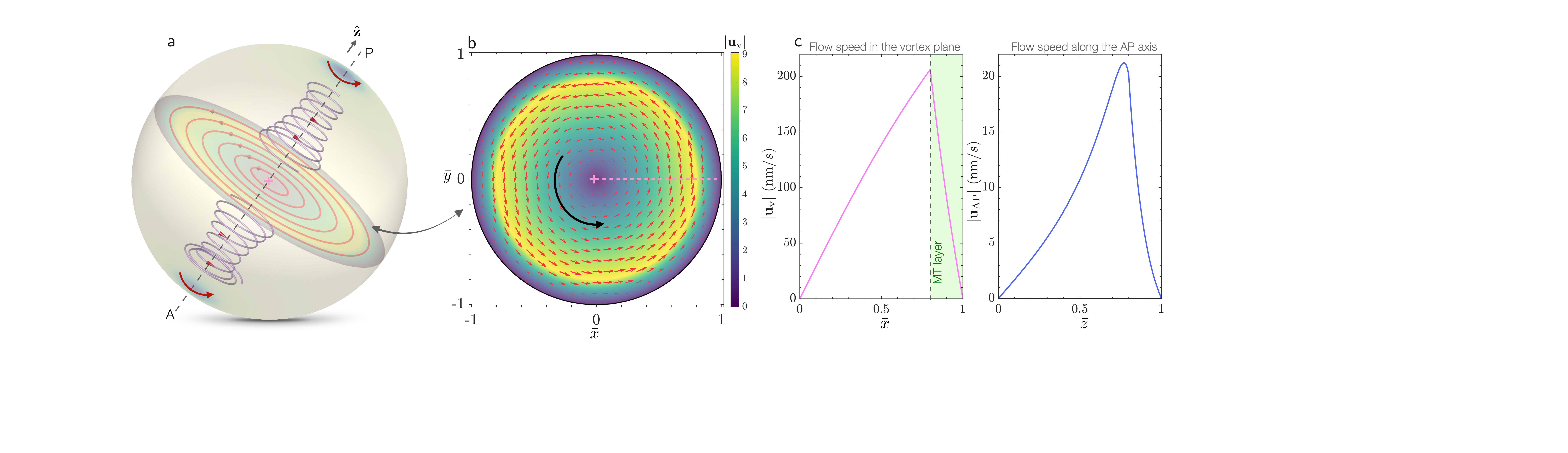}
		\caption{\textbf{\textsf{Characteristics of the twister flow.}} $\boldsymbol{\mathsf{a}}$, The vortex axis, joining the two +1 defects, is identified as the imaginary anterior-posterior (AP) axis of the abstracted cell. The vortex plane, together with the secondary swirling flow moving from the defect core to the center, is emphasized. $\boldsymbol{\mathsf{b}}$ The flow interior to the MT layer in the plane of the vortex resembles rigid body rotation. The color code is for the dimensionless in-plane velocity. $\boldsymbol{\mathsf{c}}$,  Dimensional speed as a function of the normalized distance from the vortex center in the plane of the vortex (left) and along the AP axis (right). Parameters: same as in Fig.~\ref{fig:Fig4}.}
		\label{fig:Fig5}
	\end{figure*}

	The emergent axisymmetric, steady, streaming flow is strongly vortical. This emergent flow has excellent quantitative agreements with recent particle image velocimetry experiments on late-stage oocytes \cite{dutta2023self}.  Being driven by the local polarity field, the hydrodynamics interior to the MT layer resembles a rigid-body rotation spanning the entire cell-size. In sync with our predictions from the linear theory, we find that the steady state polarity field contains a dominant contribution from the $\boldsymbol{\Phi}_1^0$ (see  Fig.~\ref{fig:Fig3}$\boldsymbol{\mathsf{c}}$(ii)) mode, responsible for the vortex flow.  The flow in the plane of the vortex as shown in  Fig.~\ref{fig:Fig5}$\boldsymbol{\mathsf{b}}$ is purely two-dimensional. In this plane, the fluid velocity increases almost linearly from the vortex core to the tips of the MT layer (see Fig.~\ref{fig:Fig5}$\boldsymbol{\mathsf{c}}$) followed by a sharp decrease in the active layer as it adapts to the no-slip boundary condition. However, this large-scale vortical flow now also features a secondary swirl component  that channels fluid inwards from the poles along the defect (or twister) axis (see Fig.~\ref{fig:Fig5}$\boldsymbol{\mathsf{a,c}}$).

	The structure of the weaker flow can be further elucidated by examining the spectral decomposition of the polarity field in its steady state. The spiraling pattern of this field around defects deviates from the pure vortex mode represented by $\mathbf{\Phi}_1^0$, leading to two key outcomes. First, a radial variation emerges in the orientation field, becoming pronounced near the defects where MTs point inwards, conforming to the geometric constraint $|\bn| = 1$. This behavior is represented by an axisymmetric mode $\mathbf{Y}_2^0$ shown in Fig.~\ref{fig:Fig3}$\boldsymbol{\mathsf{b}}$(ii). Motor proteins push fluid along the MTs pointing inwards near the poles, and incompressibility then induces a peripheral flow redirecting fluid back to the defect core. This pattern manifests as a pair of toroidal vortices around the defect axis, termed the bi-toroidal flow, highlighted in Fig.~\ref{fig:Fig3}$\boldsymbol{\mathsf{b}}$(ii).
	
	The second component to this weaker flow stems from the deviation of the projected polarity field from the equatorial lines of the sphere. Appearing as a pair of sinks near the defects, this weak deviation is described by the curl-free axisymmetric mode $\boldsymbol{\Psi}_2^0$ (shown in the SI). This arrangement of the polarity field incurs a tangential surface stress on the MT bed that induces a similar bi-toroidal flow, reinforcing the previous contribution. We emphasize that the presence of this weaker flow in the emergent state is innately nonlinear, rooted in the topological necessity to have defects and the geometric constraints set by $|\bn| = 1$.
	
	\subsection{Low-dimensional dynamics}

	Despite the complexity of our simulations and the excitation of several unstable wavenumbers (see Fig.~\ref{fig:Fig3}$\boldsymbol{\mathsf{a}}$), the streaming dynamics that surfaces is inherently low-dimensional. More importantly, its core characteristics can be distilled using select modes from the VSH basis. This insight prompts us to explore whether we can obtain coupled ordinary differential equations (ODEs) using a few salient modes that could encapsulate the essential features of this behavior. Accordingly, we propose an expansion for the polarity field as
	\begin{equation}\label{eq:expa}
		\bn = \underbracket{A(t) \mathbf{Y}_0}_{\text{radial}} + \underbracket{B(t) \boldsymbol{\Phi}_1^0}_{\text{vortex}} + \underbracket{C(t) \mathbf{Y}_2^0}_{\text{bi-toroidal}} + \underbracket{D(t) \boldsymbol{\Phi}_2^1}_{m \neq 0}.
	\end{equation}
	Here, $\mathbf{Y}_0 = -\uvc{e}_r/\sqrt{4 \pi}$ corresponds to the upright base state of the polarity field and $\{\boldsymbol{\Phi}_1^0, \mathbf{Y}_2^0, \boldsymbol{\Phi}_2^1  \}$ respectively represent the modes associated with the vortex, bi-toroidal, and non-axisymmetric flows as illustrated in Fig.~\ref{fig:Fig3}. We aim to derive the ODEs that govern the evolution of the time-dependent amplitudes $\{A(t), B(t) \cdots \}$ of each mode.  The current form of the expansion is motivated by its simplicity and ease of physical interpretation. To obtain the amplitude equations, we first approximate the interfacial traction jump as $\bar{\bff}(\by) \approx \bar{\rho}(\bar{\sigma} -\chi)\bn$, a simplification reasonable near the bifurcation. We then use Eq.~\eqref{eq:intfor} in conjunction with the no-slip boundary condition to determine the velocity field $\bu(\bx)$. Using this solution for the velocity field, we expand Eq.~\eqref{eq:polarity} up to fourth order in amplitude and use Galerkin projection to obtain the  evolution equations (see SI).
		\begin{figure*}
		\centering
		\includegraphics[width=1\textwidth]{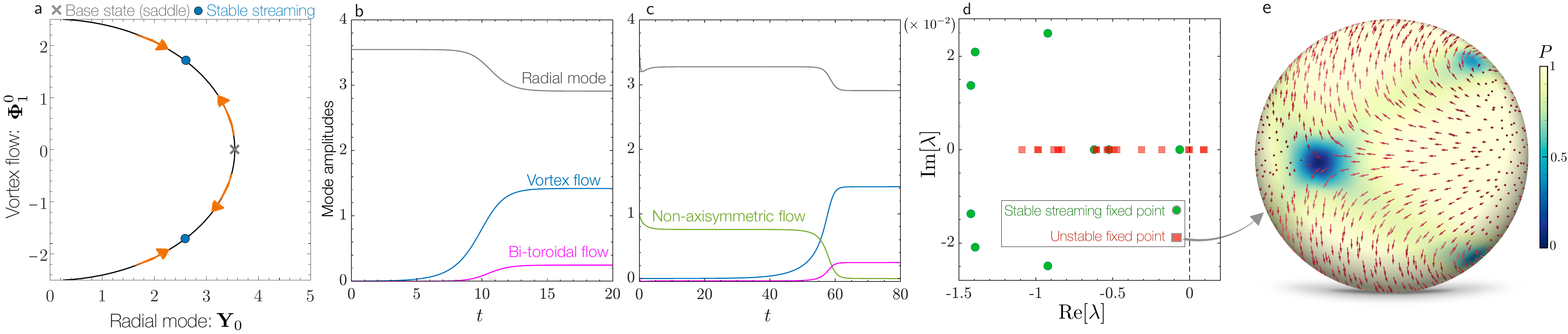}
		\caption{\textbf{\textsf{Mode dynamics, the global attractor, and unstable exact solutions.}} $\boldsymbol{\mathsf{a}}$, Phase space structure for a two-mode approximation to the streaming dynamics. The black curve represents the 1D manifold on which the dynamics is constrained to be in.  $\boldsymbol{\mathsf{b}}$, Amplitude evolution for a three-mode approximation. The bi-toroidal flow spontaneously emerges through nonlinear mode interactions. $\boldsymbol{\mathsf{c}}$, A four-mode approximation that now includes contribution from non-axisymmetric mode dynamics. The initial saturation and eventual absence of the non-axisymmetric mode hints at the possibility of unstable solutions in the problem. The emergent steady state corresponds to the stable fixed point obtained with the three mode approximation. $\boldsymbol{\mathsf{d}}$, Eigenspectra computed from a numerical linear stability analysis for the stable twister state (in green) and an exact unstable solution (in red); $\lambda$ denotes the eigenvalues. This analysis highlights that the twister is an attractor of the dynamics. The unstable branch shown in $\boldsymbol{\mathsf{e}}$ is computed numerically and features significant contributions from the non-axisymmetric mode $\boldsymbol{\Phi}_2^1$ (see Fig.~\ref{fig:Fig3}) as hinted from the reduced model.}
		\label{fig:Fig6}
	\end{figure*}
	
	To illustrate this reduced dynamics, we initiate our analysis by considering the simplest case: a two-mode  approximation. This encompasses the vortex mode and the homogeneous radial mode with $C = D = 0$.  The ODEs governing the evolution of the two relevant modes are given as
	\begin{align}
				\frac{\md A}{\md t} &=	(A^2-4\pi) \mathcal{H}(A) - \mathsf{F} A^2 B^2/4 \pi^{3/2}, \label{eq:2m1}\\
		\frac{\md B}{\md t} &=	 {A B} \mathcal{Q}(A,B), \label{eq:2m2}
	\end{align}
  \noindent where  $\mathsf{F}  = \bar{\rho} (\bar{\sigma} - \chi) (R-1)^3/R^3$ and $\mathcal{H}(A)$, $\mathcal{Q}(A,B)$ are two quadratic polynomials (see Appendix~\ref{app:2mode}).The base state $\bn = -\uvc{e}_r$ is represented by $B = 0$ and $A = \sqrt{4 \pi}$. As predicted by the linear stability analysis, this homogeneous radial state is unstable. The constraint of $|\bn| = 1$ implies that the dynamics effectively exists along a single curve which is shown in  Fig.~\ref{fig:Fig6}$\boldsymbol{\mathsf{a}}$. The dynamics on this constrained manifold reveals that the system evolves toward a stable steady state marked by a non-zero amplitude of the vortex mode, inherently tied to a streaming flow. The emergence of two stable fixed points with contrasting signs for $B$ signifies a lack of bias in the chirality of the vortex. This is further illustrated from the $B \to -B$ symmetry in Eqs.~\eqref{eq:2m1}-\eqref{eq:2m2}. The associated interior flow-field has the structure shown in Fig.~\ref{fig:Fig3}$\boldsymbol{\mathsf{c}}$ consisting of a rigid-body rotation and a shear-flow in the MT layer. 
	
	Our next approximation is motivated by  the existence of the weak bi-toroidal flow in the full computations. To understand how nonlinear interactions spontaneously give rise to this secondary flow, we consider a three mode approximation (see Eq.~70-72 of SI). This extends our previous case by including contribution from $\mathbf{Y}_2^0$ (the bi-toroidal flow), while maintaining $D = 0$. As shown in Fig.~\ref{fig:Fig6}$\boldsymbol{\mathsf{b}}$, starting from an initial condition of $C(0) = 0$, the reduced model evolves towards a steady streaming state with contributions from all three modes. Both the reduced model and computations reveal that this secondary flow consistently appears for any choice of parameters above the stability threshold. This persistence corroborates its truly nonlinear nature, embedded in the geometric constraints of the problem. 
	
	Our nonlinear simulations not only confirm the robust emergence of the axisymmetric twister in the steady state but also highlight transient non-axisymmetric flows. This motivates us to understand the interactions between axisymmetric and non-axisymmetric modes. To this end, we retain all the terms from Eq.\eqref{eq:expa} which now finally includes contribution from a non-axisymmetric mode $\boldsymbol{\Phi}_2^1$ (see Eq.~73-76 of SI). The velocity field associated with this mode is shown in Fig.~\ref{fig:Fig3}$\boldsymbol{\mathsf{b}}$(iii). To compare with our nonlinear simulations shown in Fig.\ref{fig:Fig4}$\boldsymbol{\mathsf{b}}$(ii), we choose initial conditions close to the base state and retain dominant contributions from this non-axisymmetric mode. As revealed by Fig.~\ref{fig:Fig6}$\boldsymbol{\mathsf{c}}$, the trajectory of amplitude evolution mirrors the power spectrum from the nonlinear simulations. In concert with the simulations, here we observe an initial amplitude saturation with non-axisymmetric flows. However, nonlinear mode interactions cause the vortex mode to grow. Its growth is accompanied by the decay and eventual disappearance of the non-axisymmetric component. This hints at the symmetries of the governing equations that allow the sustenance of only axisymmetric flows in steady states.
\section{Discussion}
Our simulations, combined with low-dimensional mode dynamics, firmly establish the robustness with which the cell transitions to the streaming state.  To conclusively underscore that this is indeed a stable attractor, we numerically probed the stability of the emergent fixed point. The eigenspectra depicted in Fig.~\ref{fig:Fig6}$\boldsymbol{\mathsf{d}}$ provide evidence in support of this. 
	
	Figure~\ref{fig:Fig4} revealed that starting from an almost straight configuration, the rapid onset of self-organization initiates over the fast timescale of $\tau_c$. However, the duration for complete relaxation to the streaming state shows subtle dependence on the initial conditions. Further, the quasisteady transient dynamics (see \ref{fig:Fig4}$\boldsymbol{\mathsf{b}}$(ii)) in simulations and in the reduced amplitude model, hint at the possibility of low-dimensional unstable attractors. Indeed, using a Newton-Krylov method on our governing PDEs, we numerically pinpointed one such unstable fixed point of the system, as illustrated in Fig.~\ref{fig:Fig6}$\boldsymbol{\mathsf{e}}$. Strikingly, this unstable fixed point predominantly features the non-axisymmetric, divergence-free $\boldsymbol{\Phi}_2^1$ mode, as previously alluded to by both simulations and the reduced dynamics. However, a cell is not perfect. It does not have uniformly distributed MTs, neither does it have a homogeneous activity of motor proteins. We speculate that this heterogeneity and intrinsic biochemical fluctuations potentially make the pathway towards the global attractor more robust enabling the formation of these streaming flows over a wide  range of parameters. 
	
	Cells are also not spherical. In \textit{Drosophila}, late-stage oocytes are best approximated as prolate ellipsoids with their aspect ratio being dynamic during the streaming phase as it grows in size \cite{lu2016microtubule}. This broken symmetry in the problem raises the question: how the collective dynamics and the emergent flow topologies adapt to geometry. Recent computations seem to suggest that in these egg-shaped cells, flows tend to align with the anterior-posterior (AP) axis \cite{dutta2023self} in contrast with the dynamics in a sphere, where symmetry precludes any preferential axis and the final orientation is purely set by initial conditions. How symmetry breaking leads to the loss of possible solutions and ultimately to the genesis of a single one remains unknown. Our coarse-grained PDEs provide an ideal framework to analyze such low dimensional dynamics in future.
	
	In summary, we have presented a coarse-grained active carpet theory for understanding cytoplasmic streaming inspired from the late-stage \textit{Drosophila} oocytes. This theory has the form of a boundary force field coupled to an internal Stokesian flow and is the simplest mathematical abstraction within a hierarchy of models aimed at deciphering these intracellular dynamics \cite{stein2021swirling,dutta2023self,kimura2017endoplasmic}. Through linear stability analysis, weakly nonlinear theory, and computations we have quantitatively recapitulated the key features of the emergent streaming flow. To our knowledge, here for the first time, we provide insights on the low-dimensional organization of the collective dynamics and reveal the phase-space structure leading to the twister. We posit that our minimal description of the fluid-structure interaction holds the potential for adaptation across a spectrum of biological fluid mechanics problems: from transport in bacterial carpets to ciliated propulsion \cite{darnton2004moving,lubkin2007viscoelastic,ishikawa2020stability,chakrabarti2022multiscale}.

	\appendix 
	
	\section{Vector spherical harmonics}\label{app:VSH}
	Here we provide the definition of the VSH used in the main text for analyzing solutions inside a sphere. Let $\theta \in [0, \pi]$ and $\phi \in [0,2 \pi)$ be the polar and azimuthal angles in the standard parametrization of the unit sphere. The scalar spherical harmonic $Y_\ell^m$ of degree $\ell$ and order $m$ (for $|m| \leq \ell$) is defined in terms of the associated Legendre functions $P_\ell^m$ by
			\begin{equation}
				Y_\ell^m(\theta, \phi)=\sqrt{\frac{2 \ell+1}{4 \pi}} \sqrt{\frac{(\ell-|m|) !}{(\ell+|m|) !}} P_\ell^{|m|}(\cos \theta) e^{i m \phi}.
			\end{equation}
			The VSH $\{\mathbf{Y}_\ell^m, \boldsymbol{\Psi}_\ell^m, \boldsymbol{\Phi}_\ell^m \}$ of degree $\ell$ and order $m$ (for $|m| \leq \ell$)  are defined as \cite{corona2018boundary}
			\begin{align}
				\mathbf{Y}_\ell^m 				&=-Y_{\ell}^m \uvc{e}_r, \\
				\boldsymbol{\Psi}_\ell^m   &= \nabla_\gamma Y_{\ell}^m, \\
				\boldsymbol{\Phi}_\ell^m  &= \uvc{e}_r \times \nabla_\gamma  Y_{\ell}^m,
			\end{align}
			where $\nabla_\gamma = \partial_\theta \uvc{e}_\theta + \frac{1}{\sin \theta} \partial_\phi \uvc{e}_\phi$.  Given a spectral representation of the polarity field  $\bn(\theta,\phi) = a_{\ell}^m  \mathbf{Y}_\ell^m + b_{\ell}^m  \boldsymbol{\Psi}_\ell^m  + c_{\ell}^m \boldsymbol{\Phi}_\ell^m$ the power-spectrum is defined as:
			\begin{equation}
				\begin{split}
					\mathcal{P}_\ell = \frac{1}{2 \ell +1}  \sum_m |a_{\ell}^m|^2 \langle \mathbf{Y}_\ell^m, \bar{\mathbf{Y}}_\ell^m \rangle & + |b_{\ell}^m|^2 \langle \boldsymbol{\Phi}_\ell^m, \bar{\boldsymbol{\Phi}}_\ell^m \rangle \\  & + |c_{\ell}^m|^2 \langle \boldsymbol{\Psi}_\ell^m, \bar{\boldsymbol{\Psi}}_\ell^m \rangle. 
				\end{split}
			\end{equation}
			Here we have introduced the inner-product of a vector $\bff$ defined on the unit-sphere as
			\begin{equation}
				\langle \bff, \bar{\bff} \rangle  = \int_0^{2 \pi} \int_0^\pi \bff \cdot \bar{\bff} \sin \theta \ \md \theta \md \phi,
			\end{equation}
			where ${\bar{\bff}}$ denotes the complex conjugate.
			
			\section{2-mode dynamics}\label{app:2mode}
			Equations \eqref{eq:2m1}-\eqref{eq:2m2} provide the ODEs for the 2-mode approximation. The two polynomials appearing in these equations are given as:
			\begin{align}
				\mathcal{H}(A) &= \frac{A}{\pi} - \frac{8 \pi + A^2}{8 \sqrt{\pi}},\\
				\mathcal{Q}(A,B)& = \frac{40 A \sqrt{\pi} - 5 A^2 \pi - 40 \pi^2 + 16 (5 \pi-3B^2) \mathsf{F}}{{{160 \pi^{3/2}}}},
			\end{align}
			where $ \mathsf{F}  = \bar{\rho} (\bar{\sigma} - \chi) (R-1)^3/R^3$.

			\section{Numerical method}\label{app:NM}
			In order to solve for the emergent flows, we evolve Eqs.~\eqref{eq:polarity}, \eqref{eq:jump}, and \eqref{eq:intfor} simultaneously. Given a polarity field $\bn(\by)$, Eq.~\eqref{eq:jump} provides the traction jump across the fixed interface $\partial S:= \partial D +  L \hat{\boldsymbol{\vartheta}}$. We use a 5th order accurate quadrature scheme to discretize the surface integrals of Eq.~\eqref{eq:intfor} and use quadrature by expansion to evaluate near interactions as outlined in \cite{rachh2017fast,corona2017integral}. For simulations presented here, we used $N \sim 2700$ quadrature points on the spherical surface. Using the no-slip boundary condition provides us with an integral equation of the first kind for the unknown traction $\bff^w$ on Eq.~\eqref{eq:intfor}. The resulting linear system for $\bff^w$ is solved using GMRES that typically takes $5-7$ iterations. The velocity gradient on $\partial D$ required for the evolution of the polarity field, is then determined using a second-order accurate finite difference approximation. The polarity field is evolved with a second order accurate Runge-Kutta scheme with $\Delta t \sim 5-10 \times 10^{-3}$.
			
				\acknowledgments{
				B.C is indebted to Manas Rachh for his help with computations. The authors thank Jasmin Imran Alsous, Alexandra   Jain, David Stein, Vicente Gomez Herera, David Saintillan, Dipanjan Ghosh, Shreya Biswas, Sayantan Dutta, and Reza Farhadifar for illuminating discussions and helpful feedback. MJS acknowledges support by the National Science Foundation under awards and DMR-2004469.}

			\bibliography{pnas-sample}

	\end{document}


\title{Supporting information: Cytoplasmic stirring by active carpets}
\author{Brato Chakrabarti}
\author{Stanislav Y. Shvartsman}
\author{Michael J. Shelley}
\maketitle

\section{Cytoplasm as a Stokesian fluid}
In the present problem the key dimensional  scales are: the system size $L_s \sim 300 \ \mu$m, microtubule (MT) length $L \sim 20 \ \mu$m, and   the streaming speed $u_s \sim 200$ nm/s \cite{quinlan2016cytoplasmic}. This results in a measure of shear rate as $\dot{\gamma} \sim u_s/L \sim (2-10) \times 10^{-2}$ s$^{-1}$. Experiments reveal that for this shear rate the cytoplasm behaves as a Newtonian fluid of viscosit $\mu \sim 1$ Pa.s \cite{ganguly2012cytoplasmic}. Assuming a density of water for the cytoplasmic fluid,  we estimate the Reynolds number to be:
\begin{equation}
	Re \sim 5 \times 10^{-6}.
\end{equation}
The cytoplasmic fluid is predominantly water that is why it is incompressible and the density is roughly constant. As a result, the cytoplasm behaves as a Stokesian fluid. 

\section{Details of the active carpet model}

\begin{figure}[H]
	\centering 
	\includegraphics[width=0.9\textwidth]{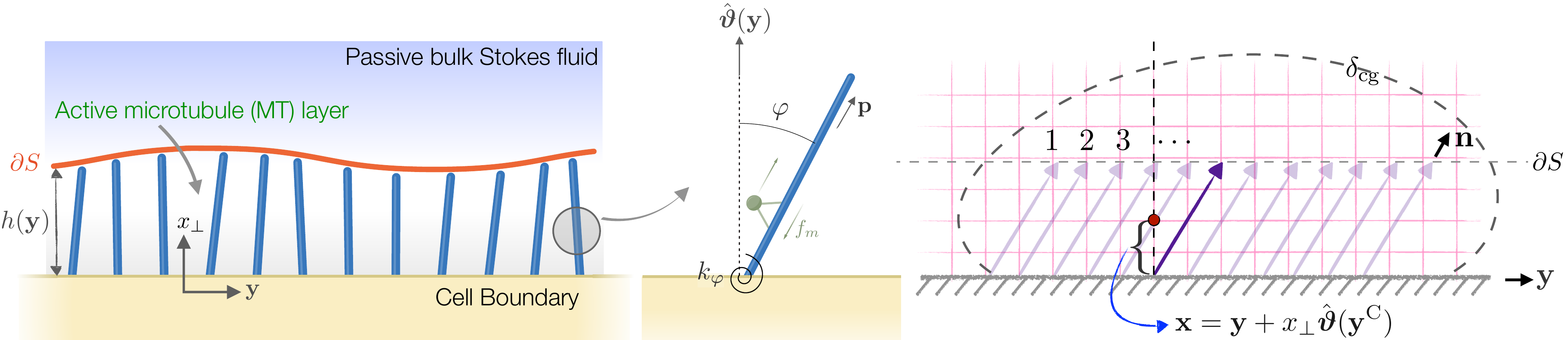}
	\caption{Left: Sketch of rods and various problem definitions. Right: Illustration of the coarse-graining volume and the relevant parameters.}
	\label{fig:Fig1}
\end{figure}

Here we provide details of the derivation of our active carpet model. We approximate the stiff MT filaments as cortically anchored rigid rods that serve as tracks for plus-end-directed Kinesin-1 motor proteins (see Fig.~\ref{fig:Fig1}$\boldsymbol{\mathsf{b}}$). The rods are equipped with a torsional spring at their anchoring point that helps them mimic the bending response of clamped flexible filaments \cite{kimura2017endoplasmic,pellicciotta2020cilia,stein2021swirling}. The dense assembly of MTs in the cell cortex motivates our formulation \cite{quinlan2016cytoplasmic}. We represent the configuration of the bed of anchored rods with a distribution function $\psi(\by,\bp,t)$. Here $\by$ is the anchoring position of the rod on the cell boundary $\partial D$ and $\bp$ is an unit vector defining its orientation. The evolution of this distribution is governed by a conservation equation
\begin{equation}\label{eq:psi}
	\partial_t \psi + \nabla_\bp \cdot (\dot{\bp} \psi) = 0,
\end{equation}
where $\nabla_\bp$ is the gradient operator on the unit sphere and $\dot{\bp}(\by,t)$ is the flux velocity in orientation. To obtain the rotational flux, consider a single MT of length $L$ and orientation $\bp$ anchored to a no-slip surface and subjected to a background flow of $\bu(\bx)$. From the slender body theory \cite{batchelor1970slender} we then have
\begin{equation}\label{eq:sbt}
	\bff(s) =  \frac{4 \pi \mu}{\log(2/\varepsilon)} \left(\bI - \frac{\bp \bp}{2}\right) \cdot \left[s \dot{\bp} - \bu(s \bp)\right] + f_m \bp,
\end{equation}
where $\bff(s)$ is the force per unit length exerted by the MT on the fluid and $f_m$ is the force per unit length exerted by a distribution of molecular motors on the MT backbone. Assuming that the background velocity $\bu(s \bp)$ can be approximated linearly at the scale of the rod, we have $\bu(s \bp) \approx s \bp \cdot \nabla \bu|_{s=0}$. The torque balance at the anchoring point yields
\begin{equation}\label{eq:tbal}
	\int_0^{L} s \bp \times \bff(s) \ \md s = k_\vphi \bT_0.
\end{equation}
The restoring torque $\bT_0$ is defined as,
\begin{equation}
	\bT_0 = -\cos ^{-1}(\mathbf{p} \cdot\hat{\boldsymbol{\vartheta}}) \frac{\mathbf{p} \times \hat{\boldsymbol{\vartheta}}}{|\mathbf{p} \times \hat{\boldsymbol{\vartheta}}|},
\end{equation}
where $\hat{\boldsymbol{\vartheta}}$ is the local surface normal to the anchoring surface. On using Eq.~\eqref{eq:sbt} in Eq.~\eqref{eq:tbal} we obtain the orientational dynamics $\dot{\bp}$ as
\begin{equation}\label{eq:pdot}
	\dot{\bp}  = (\bI - \bp \bp)  \cdot \nabla \bu\big|_{s=0} \cdot \bp + \frac{k_\vphi}{\xi_r}  \bT_0 \times \bp,
\end{equation}
where $\xi_r = 4 \pi \mu L^3/3 \log(2/\varepsilon)$ and $\varepsilon \approx 10^{-3}$ is the aspect ratio of the MT. Using Eq.~\eqref{eq:pdot} in the force balance equation we obtain the force density as 
\begin{equation}\label{eq:fsbt}
	\bff(s) = f_m \bp + \frac{4 \pi \mu s}{\log(2/\varepsilon)} \left[\frac{k_\vphi}{\xi_r} \bT_0 \times \bp - \frac{\bp \bp \bp}{2} : \nabla \bu\Big|_{s=0}\right].
\end{equation}
We now emphasize, that in our problem $\bu(\bx)$ is the mean-field fluid flow generated by the presence of other MTs in the cell periphery. Thus the above force density that drives the flow needs to be determined self-consistently along with the solution to the incompressible Stokes equation. The flow induced due to one MT can be obtained as the solution to the singularly forced Stokes equation as 
\begin{equation}
	-\nabla q + \mu \Delta \bu + \int_0^{L} \bff(s,\bp) \delta(\bx - \by - s \bp) \ \md s = \mathbf{0}.
\end{equation}
We now aim to compute a coarse-grained force generated by a distribution of such anchored MTs. To this end, we focus on a very small coarse-graining volume of size $\delta_\text{cg}(\by) \sim \ell_\text{cg}^3$ centered around a surface point $(\by^\text{C})$. We now use the following assumptions: 
\begin{itemize}
	\item We assume that in the coarse-graining box all the rods are sharply aligned yielding
	\begin{equation}
		\psi(\by,\bp,t) = c_0 \delta[\bp - \bn(\by,t)],
	\end{equation}
	where $\bn(\by)$ is the polarity  and $c_0$ is the surface density of the MT bed.
	\item At the scale of the coarse-graining volume $\ell_{\text{cg}}$, the curvature of the cell surface is negligible and can be approximated as a flat wall. 
	\item We also assume that there is a ordering of length scales in the problem as
	\begin{equation}
		\delta \ll \ell_\text{cg} \ll \Lambda \ll L_s,
	\end{equation}
	where $\delta$ is the spacing between the anchored filaments and $\Lambda$ is a characterisitic length scale over which one observes the fluctuations in any coarse-grained quantity. This formally establishes the underlying long-wavelength approximation to our theory.
\end{itemize}
The driving at the scale of a coarse-graining element can then be written as
\begin{equation}
	-\nabla q + \mu \Delta \bu + \sum_{\alpha \in \delta_{\text{cg}}} \int_0^{L} \bff(s^\alpha,\bp^\alpha) \delta(\bx-\by^\alpha - s^\alpha \bp^\alpha) \md s^\alpha  = \mathbf{0}, \ \ \forall \bx \in  \delta_{\text{cg}},
\end{equation}
where $\alpha = 1,2, \cdots$ is the index of filaments contained within the coarse-graining volume (see Fig.~\ref{fig:Fig1}). Due to our sharply aligned approximation, the MTs contained in this volume are oriented along $\bn(\by^\text{C})$ for all $\by^\alpha \in \delta_\text{cg}$. 
Realizing that $\bx = \by + x^\perp \hat{\boldsymbol{\vartheta}}(\by^\text{C}) \equiv (\by,x^\perp)$ we can write down the forcing term in the above equation as
\begin{equation}
	\mathbf{F}(\by^\text{C},x^\perp) = \sum_{\alpha \in \delta_{\text{cg}}}  \int_0^{L}  \bff(s^\alpha,\bp^\alpha) \delta(\bx-\by^\alpha - s^\alpha \bp^\alpha) \md s^\alpha = \sum_{\alpha \in \delta_{\text{cg}}}  \int_0^{L} \bff(s,\bn) \delta[\by-\by^\alpha] \ \delta\left[x^\perp - s \bn(\by^\text{C}) \cdot \hat{\boldsymbol{\vartheta}} (\by^\text{C})\right] \md s.
\end{equation}
Here $x^\perp$  is the coordinate normal to the cell boundary and in the final expression we have dropped the superscript $\alpha$ as all the MTs in the coarse-graining volume behave identically. We now introduce a change of variables $s \bn(\by^\text{C}) \cdot \hat{\boldsymbol{\vartheta}}(\by^\text{C}) \to \tilde{s}$ to re-arrange the above expression and obtain
\begin{equation}
	\mathbf{F}(\by^\text{C},x^\perp)   = \sum_{\alpha \in \delta_{\text{cg}}}  \frac{1}{\bn \cdot \hat{\boldsymbol{\vartheta}}} \int_0^{\tilde{l}_0}\bff(\tilde{s}/\bn \cdot \hat{\boldsymbol{\vartheta}},\bn) \delta[\by-\by^\alpha] \ \delta\left[x^\perp - \tilde{s}\right] \md \tilde{s} = \sum_{\alpha \in \delta_{\text{cg}}} \frac{ \bff(x^\perp/\bn \cdot \hat{\boldsymbol{\vartheta}},\bn) }{\bn \cdot \hat{\boldsymbol{\vartheta}}}  \delta(\by - \by^\alpha) = \frac{ \bff(x^\perp/\bn \cdot \hat{\boldsymbol{\vartheta}},\bn) }{\bn \cdot \hat{\boldsymbol{\vartheta}}} c_0.
\end{equation}
Using the expression for $\bff(s)$ from Eq.~\eqref{eq:fsbt} we obtain the driving force per unit volume as
\begin{equation}
	\mathbf{F}(\by^\text{C},x^\perp) = c_0 \left\{\frac{4 \pi \mu x^\perp}{\log(2/\varepsilon)(\bn \cdot \hat{\boldsymbol{\vartheta}})^2} \left[\frac{k_\vphi}{\xi_r} \bT_0 \times \bn - \frac{\bn \bn \bn}{2}  : \nabla \bu\Big|_{\partial D}  \right] + \frac{f_m}{{\bn \cdot \hat{\boldsymbol{\vartheta}}}} \bn \right\}.
\end{equation}
In our theory, we approximate the action of this volume forcing by a traction jump of $\bar{\bff}(\by)$ across the interface $\partial S$ delianated by the tips of the MT. This traction jump is then computed as
\begin{equation}
	\bar{\bff}(\by) = \int_0^{h(\by)} \mathbf{F}(\bx) \ \md x^\perp,
\end{equation} 
$h(\by)$ is the thickness of the local MT layer. Noting that $h(\by) = L \bn \cdot \hat{\boldsymbol{\vartheta}}$ we obtain
\begin{equation}
	\bar{\bff}(\by,t) =  c_0 L f_m \mathbf{n} + \frac{2 \pi \mu c_0  L^2}{\log(2/\varepsilon)}\left[\frac{k_\vphi}{\xi_r} \mathbf{T}_0 \times \mathbf{n}-\frac{\mathbf{n n n}}{2}: \nabla \mathbf{u}\Big|_{\partial D} \right]. 
\end{equation}
This completes our formulation for the active carpet theory. To non-dimensionalize the governing equations we use the following scales: length $\sim L$, time $\sim \xi_r/k_\vphi \equiv \tau_r$ (relaxation time of a single MT) and force $\sim \mu L^2/\tau_r$. This leads to the following dimensionless quantities:
\begin{alignat}{3}
	&1. \ \text{Surface density:} \ &\bar{\rho} &= c_0 L^2 \equiv \frac{N_{\text{MT}}}{A_s} L^2 \sim \frac{\delta^2}{d^2} \\
	&2. \ \text{Activity:}  \ &\bar{\sigma} &= \frac{4 \pi  f_m \ell^2}{3 \log(2/\varepsilon) k_\varphi} \\
	&3. \ \text{System size:} \ &\mathcal{R}&= \frac{L_s}{L}  \\
	&4. \ \text{Geometric parameter:} \ &\chi &= \frac{2 \pi}{\log(2/\varepsilon)} \\
\end{alignat}
Here $N_{\text{MT}}$ is the number of MTs on the cell boundary and $A_s$ is the surface area of the cell. The dimensionless traction jump is then given as
\begin{equation}\label{eq:trac}
	\bar{\bff}(\by) = \bar{\rho} \bar{\sigma} \bn + {\bar{\rho} \chi} \left[\mathbf{T}_0 \times \mathbf{n}-\frac{\mathbf{n n n}}{2}: \nabla \mathbf{u}\right] .
\end{equation}
To conclude, we provide a summary of the dimensionless governing equations:
\begin{align}
	-\nabla q +  \Delta \bu = \mathbf{0},  \ \ \nabla \cdot \bu  = 0, \label{eq:homst}& \\ 
	\bu(\by) = \mathbf{0}, \ \ \text{ on } \partial D &, \\
	\llbracket \bu \rrbracket = \mathbf{0}, \ \ \llbracket \boldsymbol{\sigma}  \cdot \hat{\boldsymbol{\vartheta}} \rrbracket (\by,t) = -\bar{\bff}(\by,t), \ \ \text{ on } \partial S, & \label{eq:stokeappx}
\end{align}
where $\llbracket a \rrbracket = a|_{\partial S^+} - a|_{\partial S^-}$ denotes the jump of any variable across the interface $\partial S$. The above set of equation completely specifies the hydrodynamics of our problem. They are coupled to the evolution of the polarity field as
\begin{equation}\label{eq:polarity}
	\partial_t \bn(\by,t)  = (\bI - \bn \bn) \cdot \nabla \bu \Big|_{\partial D} \cdot \bn +  \bT_0 \times \bn. 
\end{equation}

\section{Parameter selction}

The estimate of the dimensional parameters in our problem is provided below
\begin{table}[H]
	\centering
	\begin{tabular}{c|c|c}\hline
		Variable             	      & Value                      & Comment \\ \hline
		MT length   	 	 	   & $L \sim 20 \ \mu$m \cite{howard2002mechanics}         &         		\\
		Bending rigidity   	    & $B \sim 2 \times 10^{-23}$ N m$^2$   \cite{howard2002mechanics}       &        		\\    
		MT aspect ratio        & $\varepsilon =  10^{-3}$  \cite{howard2002mechanics}     &  \\
		Fluid viscosity           & $\mu \sim 1000 \times \mu_{\text{water}} = 1$ Pa.s \cite{ganguly2012cytoplasmic} &  \\
		Sphere radius           & $L_s \sim 5 L$ &  \\ 
		Rotational friction     & $\xi_r = 4 \pi \mu L^3/3 \log(2/\varepsilon) \sim 4.4 \times 10^{-15}$ N.m.s & \\
		Number of MTs        & ${N}_{\text{MT}} \sim 1-4 \times 10^3$ \cite{dutta2023self} & \\
		{Motor force density} & $f_m \sim \left(1- 10\right) \times 10^{-7}$N/m & A single kinesin-$1$ motor applies $ 2 \ $pN force and  $1-10$ motors/MT \\
		{Spring constant} & $k_\varphi (4-5) \sim B/L \sim \left(4-5\right) 10^{-18}$ N.m & Obtained from a callibration process as outlined in the main text  \\ \hline
	\end{tabular}
	\caption{Dimensional values of various physical parameters}
\end{table}
Taken together, this means that  the single MT relaxation time is $\tau_r \sim \xi_r/k_\vphi \sim  \times 10^3$ s and the velocity scale is given as $U \sim L/\tau_r \sim 20$ nm/s.

\section{Streaming speed in half-space}
Here we study the emergent streaming speed in the half-space problem. We are interested in the homogeneous state where all the MTs are aligned with an angle $\vphi_s$ with the $\uvc{z}$ direction as highlighted in the main text. The polarity field is given as $\bn = \cos \vphi_s\uvc{z} + \sin \vphi_s\uvc{x}$ and the restoring torque is $\bT_0 = -\vphi \uvc{y}$. The solution to the homogeneous Stokes equation inside the MT layer is given as $\bu^\text{M} = \dot{\gamma}z \uvc{x}$, where $\dot{\gamma}$ is the stead-state shear rate. The solution above the MT layer is a constant velocity profile given as $\bu^\text{B} = \dot{\gamma} \cos \vphi_s \uvc{x}$.
The associated dimensionless traction jump is given as:
\begin{align}
	\bar{\bff} &=  \bar{\rho} \chi \left[\bT_0 \times \bn - \frac{1}{2} \bn \bn \bn : \dot{\gamma} \uvc{x} \uvc{z}\right] + \bar{\rho} \bar{\sigma} \bn = \rho \chi \left[-\vphi \left(\cos \vphi \uvc{x} - \sin \vphi \uvc{z} \right) - \frac{\dot{\gamma} }{2} \cos \vphi \sin \vphi \bn\right] + \bar{\rho} \bar{\sigma} \bn. 
\end{align}
The tangential stress jump at the interface then yields $-\dot{\gamma} =  -\bar{\bff} \cdot \uvc{x}$. Using the above definitions for the  forces we obtain the following equation for the shear rate
\begin{equation}
	\dot{\gamma} = \bar{\rho} \bar{\sigma} \sin \vphi_s  - \bar{\rho} \left[\frac{\dot{\gamma}}{2} \cos \vphi_s \sin^2\vphi_s + \chi \vphi \cos \vphi_s \right].
\end{equation}
This allows us to solve for the unknown shear rate and we obtain:
\begin{equation}
	\dot{\gamma} = \frac{2 \bar{\rho}  (\bar{\sigma}  \sin \vphi_s - \chi  \vphi_s  \cos \vphi_s )}{2+  \bar{\rho} \chi  \sin^2 \vphi_s \cos \vphi_s }.
\end{equation}
We also note that the evolution equation of the angle $\vphi$ is given as: $\partial_t \vphi = \dot{\gamma} \cos^2 \vphi_s - \vphi_s$. At steady state $\partial_t \vphi_s = 0 \implies \dot{\gamma} = \vphi_s/\cos^2\vphi_s$. For a self-consistent solution, we must have
\begin{equation}
	\frac{2 \bar{\rho}  (\bar{\sigma}  \sin \vphi_s - \chi  \vphi_s  \cos \vphi_s )}{2+  \bar{\rho} \chi  \sin^2 \vphi_s \cos \vphi_s }= \frac{\vphi_s}{\cos^2 \vphi_s}.
\end{equation}
The above equation sheds light on a few aspects of the problem that we list below.
\begin{itemize}
	\item  The trivial solution to the above equation is $\vphi_s = 0$ which is the homogeneous based state of the problem with $\bu = \mathbf{0}$.
	\item A non-zero solution of $\vphi_s$  exists beyond the streaming instability. 
	\item The streaming speed for this solution is:
	\begin{equation}
		u_s = \dot{\gamma} z\Big|_{z=\cos \vphi_s} =  \frac{2 \bar{\rho}  (\bar{\sigma}  \sin \vphi_s - \chi  \vphi_s  \cos \vphi_s)}{2+  \bar{\rho} \chi  \sin^2 \vphi_s\cos \vphi_s} \cos \vphi_s.
	\end{equation}
\end{itemize}
In general, the above solution needs to be determined numerically. However, for$\bar{\rho} \to \infty$ we can obtain it analytically. In this limit the MTs are highly aligned with the $x$ axis. We define the change of variables $\vphi_s^* = \pi/2-\vphi_s$. With this, the self-consistency condition reduces to:
\begin{equation}
	\frac{2 (\bar{\sigma} \cos \vphi_s^* -\chi  \left(\frac{\pi}{2}-\vphi_s^*\right)  \sin \vphi_s^* )}{\chi   \cos^2 \vphi_s^* \sin \vphi_s^*}= \frac{\frac{\pi}{2}-\vphi_s^*}{\sin^2 \vphi_s^*}.
\end{equation}
Expanding the above equation around $\vphi_s^* = 0$ leads to the following equation:
\begin{equation}
	\frac{2 \bar{\sigma}}{\chi \vphi_s^*} - \pi + \mathcal{O}({\vphi_s^*})= \frac{\frac{\pi}{2}-\vphi_s^*}{\vphi_s^{*2}}
\end{equation}
The above equation can be solved for $\vphi_s^*$ and we can obtain the learge density asymotote as
\begin{equation}
	u_s\Big|_{\bar{\rho} \to \infty} = \frac{\sqrt{ \left[2 \bar{\sigma}  + \chi \right]^2- 2 \pi ^2 \chi ^2}+ 2 \bar{\sigma}  -  \chi }{2
		\chi }.
\end{equation}
As highlighted in the main text, the above expression is independent of the density $\bar{\rho}$ and depends only on the activity. 

\section{Linear stability analysis in a disk}
We now consider the stability analysis inside a disk of dimensionless radius $R > 1$. The base state of the problem is given as $\bn_0 = -\uvc{e}_r$, where $\uvc{e}_r$ is the unit vector pointing radially outwards.  We use the following Ansatz for any variable $f = \sum_m f_m(r) e^{\mi m \theta}$, where $\theta \in [0,2\pi)$ and $m \in \mathbb{Z}$. The Stokes equations in cylindrical coordinates with the above Ansatz for velocity and pressure can be written as:
\begin{alignat}{3}
	&\text{Continuity:} \hspace*{1mm} &\frac{1}{r} \frac{\md}{\md r}\left(r u_r\right)+\frac{\mi m}{r} u_\theta &=0, \\ 
	&\text{$r$-momentum:} \hspace*{1mm} &-\frac{\md p}{\md r}+ \left[\Delta2 u_r-\frac{u_r}{r^2}-\frac{2 \mi m}{r^2} u_\theta\right] &=0, \\ 
	&\text{$\theta$-momentum:} \hspace*{1mm} &-\frac{\mi m}{r} p + \left[\Delta u_\theta-\frac{u_\theta}{r^2}+\frac{2 \mi m}{r^2} u_r \right] &= 0, \\
	&\text{Laplace equation:}  \hspace*{1mm} &\left( \frac{\md^2}{\md r^2}+\frac{1}{r} \frac{\md}{\md r}-\frac{m^2}{r^2}\right) p &= 0.\\ 
\end{alignat}
We now note that the perturbed field of the fibers have the following form:
\begin{equation}
	\bn = -\uvc{e}_r + \underbrace{\epsilon n_m(r) e^{\mi m \theta}}_{n_\theta} \uvc{e}_\theta.
\end{equation}
With the above Ansatz we have: $\bT_0 = -\bn \cdot \uvc{e}_\theta$ and $\bT_0 \times \bn = -n_\theta \uvc{e}_\theta$. The 
stability of the bed is dictated by the following equation:
\begin{equation}
	\partial_t n_\theta = -\partial_r u_\theta\big|_{r=R}-n_\theta.
\end{equation}
This allows us to solve for the growth rate for each  azimuthal wavenumber $m$ and we obtain:
\begin{equation}
	\lambda(m) = \frac{\bar{\rho} (\bar{\sigma}-\chi)}{2}\left[1-\frac{1}{R}\right]^m \left[\frac{m}{R^2} + 2 \left(1-\frac{1}{R}\right)^2 - 2 \frac{m}{R}\right] -1.
\end{equation}
First we note that the axisymmetric mode corresponding to $m=0$ has the largest growth rate. In that case, the dispersion relation simplifies to
\begin{equation}
	\lambda(m=0) = {\bar{\rho} (\bar{\sigma}-\chi)}\left(1-\frac{1}{R}\right)^2 -1.
\end{equation}
This closely resembles our dispersion relation for the half-space with a small modulation coming from the disk radius $R$. In order to further comapre our results with the half-space problem we take the limit ${m,R} \to \infty$ with $m/R \to k$ remaining finite. Using the relation
\begin{equation}
	\lim_{m,R \to \infty, m/R \to k} \left[1-\frac{1}{R}\right]^m = \lim_{m\to \infty} \left[1-\frac{k}{m}\right]^m = e^{-k},
\end{equation} 
we find
\begin{equation}
	\lambda_k = \bar{\rho} (\bar{\sigma}-\chi)e^{-k}(1-k)-1, 
\end{equation}
identical to that obtained in the main text.

\section{Linear stability for sphere}
The base state of the problem is $\bn_0 = -\uvc{e}_r$, where $\uvc{e}_r$ is the unit-outward normal to the sphere. To probe the stability of this state, we consider the evolution of a slightly perturbed polarity field $\bn = -\uvc{e}_r + \epsilon \tilde{\bn}(\theta,\phi) e^{\lambda t} \ (\epsilon \ll 1)$, where $\tilde{\bn}$ lies in the tangent-space. Realizing that the associated flow $\bu \sim \mathcal{O}(\epsilon)$, the evolution of the perturbation from Eq.~\eqref{eq:polarity} at the linear-order is given as
\begin{equation}\label{eq:linsp}
	\lambda \tilde{\bn} = -\partial_r \bu \big|_{r=R} - \tilde{\bn}.
\end{equation}
The associated linearized traction jump is given as
\begin{equation}\label{eq:lintheta}
	\bar{\bff} = \bar{\rho} (\bar{\sigma} - \chi) \tilde{\bn}.
\end{equation}
To obtain the disturbance flow $\bar{\bu}$ we make use of the integral representation of the Stokes equation as outlined in the main text. This is given as,
\begin{equation}
	\bu(\bx) = - \mathcal{S}[\bff^w](\bx) \Big|_{\partial D} + \mathcal{S}[\bar{\bff}](\bx)\Big|_{\partial S},
\end{equation}
where $\bff^w$ is the wall traction. Here $\mathcal{S}[\boldsymbol{\sigma}]$ is the single layer operator defined as
\begin{equation}
	\mathcal{S}[\boldsymbol{\sigma}](\boldsymbol{x})\Big|_{\Gamma} = \frac{1}{8 \pi} \int_{\Gamma} \boldsymbol{\sigma} (\bx_0) \cdot \mathcal{G}(\bx,\bx_0) d A(\bx_0), 
\end{equation}
where $\mathcal{G}(\bx,\bx_0)$  is the free-space Green's function for the Stokes equation as defined earlier. We now expand the polarity field and the unknown traction $\bff^w$ in the basis of the vector spherical harmonics (VSH) as 
\begin{align}
	\tilde{\bn}(\theta,\phi) &= \sum_{\ell \geq 1, |m| \leq \ell} b_{\ell}^m  \boldsymbol{\Psi}_\ell^m  + c_{\ell}^m \boldsymbol{\Phi}_\ell^m, \label{eq:nvsh} \\
	\bff^w &= d_{\ell}^m \mathbf{Y}_\ell^m + e_{\ell}^m  \boldsymbol{\Psi}_{\ell}^m + f_{\ell}^m  \boldsymbol{\Phi}_{\ell}^m \label{eq:walltrac}.
\end{align}
Here we have introduced the set $\{\mathbf{Y}_\ell^m, \boldsymbol{\Psi}_\ell^m, \boldsymbol{\Phi}_\ell^m \}$, which are three orthogonal VSH of degree $\ell$ and order $m$ (for $|m| \leq \ell$) with $\ell$ being the polar and $m$ the azimuthal wavenumber \cite{hill1954theory}. Given a vector field represented in this basis, $\mathbf{Y}_\ell^m$ delineates its radial variation, $\boldsymbol{\Psi}_\ell^m$ accounts for the curl-free  and $ \boldsymbol{\Phi}_\ell^m$  the divergence free component. For the sake of completeness we provide their respective definitions here:
\begin{align}
	\mathbf{Y}_\ell^m 				&=-Y_{\ell}^m \uvc{e}_r, \\
	\boldsymbol{\Psi}_\ell^m   &= \nabla_\gamma Y_{\ell}^m, \\
	\boldsymbol{\Phi}_\ell^m  &= \uvc{e}_r \times \nabla_\gamma  Y_{\ell}^m,
\end{align}
where $Y_\ell^m$ is the scalar spherical harmonic. The Ansatz for $\tilde{\bn}$ in Eq.~\eqref{eq:nvsh} ensures that it is in the tangent space. The unknown traction $\bff^w$ is obtained by using the no-slip condition $\bu(\bx) = \mathbf{0}$ on $\partial D$. Now we use the fact that the the VSHs are the eigenfunctions of the single layer operator defined on the sphere \cite{schmitz1982creeping,corona2018boundary,kim2013microhydrodynamics}. Given a sphere of radius $R$, we have the following relationships for the present choice of the VSH basis: 
\newpage
\begin{equation*}
	\text{Solution for $r \geq R$:}
\end{equation*}
\begin{footnotesize}
	\begin{align}
		\mathcal{S}[\mathbf{Y}_\ell^m] &= \frac{r^{-2 - \ell} R^{1 + \ell}}{2\ell+1}\left\{\left[\frac{\ell(1 + \ell)  (2\ell + 3) r^2 + \ell(1 + \ell) (2\ell - 1) R^2}{8\ell(1 + \ell) - 6}\right]  \mathbf{Y}_\ell^m + \left[\frac{(6 + \ell - 2 \ell^2) r^2 + \ell (2\ell - 1) R^2}{8\ell(1 + \ell) - 6}\right] \boldsymbol{\Psi}_\ell^m\right\}, \\
		\mathcal{S}[\boldsymbol{\Psi}_\ell^m] &= \frac{\ell+1}{2\ell+1}r^{-2 - \ell} R^{1 + \ell} \left\{\left[\frac{(\ell + \ell^2)(2\ell + 3) r^2 - (3 \ell + \ell^2)(2\ell - 1) R^2}{8\ell(1 + \ell) - 6}\right]  \mathbf{Y}_\ell^m + \left[\frac{{(2 - \ell)(1 + \ell)(2\ell + 3) r^2 + \ell(3 + \ell)(2\ell - 1) R^2}}{2 \ell + 8 \ell^2(2 + \ell) -6 }
		\right] \boldsymbol{\Psi}_\ell^m\right\}, \\
		\mathcal{S}[\boldsymbol{\Phi}_\ell^m] &= \frac{1}{2\ell + 1} \frac{R^{\ell + 2}}{r^{\ell + 1}} \boldsymbol{\Phi}_\ell^m.
	\end{align}
\end{footnotesize}
\begin{equation*}
	\text{Solution for $r \leq R$:}
\end{equation*}
\begin{footnotesize}
	\begin{align}
		\mathcal{S}[\mathbf{Y}_\ell^m] &= \frac{r^{-1 + \ell} R^{-\ell} }{2 \ell+1}\left\{\left[\frac{\ell(1 + \ell) (2\ell + 3) R^2 - \ell(1 + \ell) (2\ell - 1) r^2}{8\ell(1 + \ell) - 6}\right] \mathbf{Y}_\ell^m + \left[\frac{- (3 + \ell)(2\ell - 1) r^2 + (1 + \ell)(2\ell + 3) R^2}{8\ell(1 + \ell) - 6}\right] \boldsymbol{\Psi}_\ell^m\right\}, \\
		\mathcal{S}[\boldsymbol{\Psi}_\ell^m] &= \frac{\ell+1}{2 \ell+1}  r^{-1 + \ell} R^{-\ell} \left\{\left[\frac{6\ell R^2 + \ell^2 (2\ell - 1) (r - R) (r + R)}{8\ell(1 + \ell) - 6}\right] \mathbf{Y}_\ell^m + \left[\frac{\ell(3 + \ell)(2\ell - 1) r^2 - (\ell -2)(1 + \ell)(2\ell + 3) R^2 }{2 \ell + 8 \ell^2(2 + \ell) -6 }\right] \boldsymbol{\Psi}_\ell^m \right\}, \\
		\mathcal{S}[\boldsymbol{\Phi}_\ell^m] &=\frac{1}{2\ell + 1} \frac{r^{\ell}}{R^{\ell - 1}} \boldsymbol{\Phi}_\ell^m.
	\end{align}
\end{footnotesize}
Using the above relations in conjunction with the no-slip boundary conditions we can now express the coefficients of Eq.~\eqref{eq:walltrac} in terms of $\tilde{\bn}$. They are given as follows:
\begin{align}
	d_{\ell}^m &=\frac{1}{2} (R-1)^{1 + \ell} R^{-3 - \ell} \left[(3 + \ell) \left(a_\ell^m + b_\ell^m + b_\ell^m \ell\right) R^2  -(1 + \ell) \left(a_\ell^m + b_\ell^m (3 + \ell)\right) (R-1)^2 \right], \\
	e_\ell^m &= \frac{1}{2} (R-1)^{1 + \ell} R^{-3 - \ell} \left[ \left(a_\ell^m + b_\ell^m (3 + \ell)\right) (R-1)^2 - \left(a_\ell^m + b_\ell^m + b_\ell^m \ell\right) R^2 \right], \\
	f_\ell^m &= (R-1)^{2 + \ell} R^{-2 - \ell}.
\end{align}
Using the above solution we can now rewrite Eq.~\eqref{eq:lintheta} in the following form
\begin{equation}
	(\lambda+1) \left[b_{\ell}^m  \boldsymbol{\Psi}_\ell^m  + c_{\ell}^m \boldsymbol{\Phi}_\ell^m\right] = \partial_r \left[-\mathcal{S}[\bff^w]\Big|_{\partial D \to \partial D}+ \mathcal{S}[\bar{\bff}]\Big|_{\partial S \to \partial D}\right]\Big|_{r= R}.
\end{equation}
where we have used the Ansatz for $\tilde{\bn}$. In the notation $\partial S \to \partial D$, the first term highlights the surface on which the traction is defined and the second term highlights highlights the surface on which it is evaluated. This provides us with an eigenvalue problem for the growth rate $\lambda$ for a discrete pair of $\{\ell,m\}$. For a given choice of $\{\ell,m\}$ we find two eigenvalues $\lambda_{1,2}$. The dominant eigenvalue is always associated with the coefficient $c_{\ell}^m$ and thus corresponds to the divergence free mode $\boldsymbol{\Phi}_{\ell  m}$. This dominant eigenvalue is given as
\begin{equation}\label{eq:lambda}
	\lambda(\ell) = -1 +  \left(\frac{R-1}{R}\right)^{\ell+2} \bar{\rho} (\bar{\sigma}-\chi),
\end{equation}
with the most unstable mode being $\ell = 1$. The associated eigenmode for velocity is computed from the integral representation. This is constructed by stitching two pieces: (a) an solution interior to $\partial S$ and (b) a solution in the thing gap between $\partial S$ and $\partial D$. 

\section{Contributions to the bi-toroidal flow}
\begin{figure}[H]
	\centering 
	\includegraphics[width=0.6\textwidth]{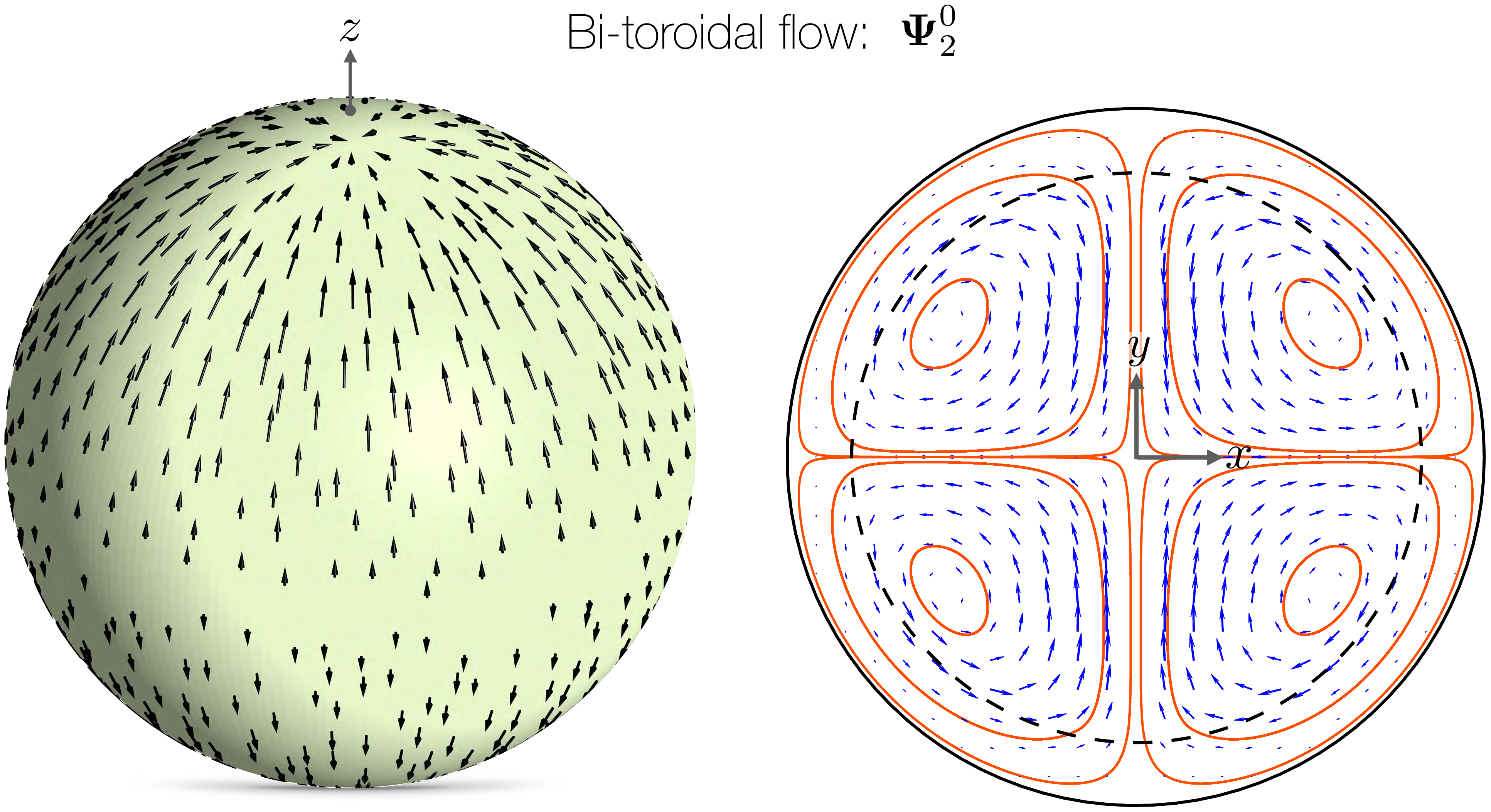}
	\caption{Surface polarity field due to the axisymmetric curl-free mode (on left). The associated cross-sectional velcity field on the right.}
\end{figure}
In the main text we have highlighted that there are two contributions to the weak bi-toroidal flow in the emergenst state of the twister. The main text pointed out the mode $\mathbf{Y}_2^0$ that emerges from variation of the polarity field in the radial direction. Here we highlight the second contribution to the weaker flow appearing due to variations in the surface polarity field.

\section{Weakly nonlinear theory and ODEs}

For our weakly nonlinear theory we expanded the polarity field as follows:
\begin{equation}\label{eq:expa}
	\bn = \underbracket{A(t) \mathbf{Y}_0}_{\text{radial}} + \underbracket{B(t) \boldsymbol{\Phi}_1^0}_{\text{vortex}} + \underbracket{C(t) \mathbf{Y}_2^0}_{\text{bi-toroidal}} + \underbracket{D(t) \boldsymbol{\Phi}_2^1}_{m \neq 0}.
\end{equation}
We recall from Eq.~\eqref{eq:trac} that the traction jump across $\partial S$ is given as
\begin{equation}
	\bar{\bff}(\by) = \underbracket{\bar{\rho} \bar{\sigma} \bn}_{\text{linear in } \bn} + \underbracket{{\bar{\rho} \chi} \left[\mathbf{T}_0 \times \mathbf{n}-\frac{\mathbf{n n n}}{2}: \nabla \mathbf{u}\right]}_{\text{nonlinear in } \bn} .
\end{equation}
For our analysis we retain only the linear bit in the traction, an approximation  fairly accurate close to the bifurcation and write it as:
\begin{equation}
	\bar{\bff}(\by) \approx \bar{\rho}(\bar{\sigma}-\chi)\bn. 
\end{equation}
We retain all the nonlinearities in Eq.~\eqref{eq:polarity} for the evolution of the polarity field. To derive amplitude equations we make use of the no-slip boundary condition to compute the wall-tractions $\bff^w$. This allows us to compute the velocity gradient on $\partial D$ necessary for the evolution of the polarity field. By projecting on the appropriate basis we then obtain the amplitude equations. Here we provide the details of the ODEs derived using \textit{Mathematica} that we solved in the main text. For the sake of notation we also introduce $\mathcal{F} = \bar{\rho}(\bar{\sigma}-\chi)$.

\subsection{2 mode approximation}
\begin{align}
	\frac{\md A}{\md t} &=	-A + \frac{A^3}{4 \pi} - \frac{A^4}{32 \sqrt{\pi}} + \pi^{3/2} - \frac{A^2 \left[(\pi^2 R^3 + 2 B^2 (-1 + R)^3 \mathcal{F}\right]}{8 \pi^{3/2} R^3}, \\
	\frac{\md B}{\md t} &=	\frac{A B}{{160 \pi^{3/2}}} \left[40 A \sqrt{\pi} - 5 A^2 \pi - 40 \pi^2 + \frac{16 (-3 B^2 + 5 \pi) (-1 + R)^3 \mathcal{F}}{R^3}\right].
\end{align}

\subsection{3 mode approximation}
\begin{scriptsize}
	\begin{align}
		\begin{split}
			\frac{\md A}{\md t} =&	-A + \frac{7 A^3 + 21 A C^2 + 2 \sqrt{5} C^3}{28 \pi} - \frac{7 A^4 + 42 A^2 C^2 + 8 \sqrt{5} A C^3 + 15 C^4}{224 \sqrt{\pi}} - \frac{1}{8} (A^2 + C^2) \sqrt{\pi} + \pi^{3/2} \\
			&- \frac{B^2 (35 A^2 - 14 \sqrt{5} A C + 25 C^2) (-1 + R)^3 \mathcal{F}}{140 \pi^{3/2} R^3}, 
		\end{split}\\
		\begin{split}
			\frac{\md B}{\md t} = &\frac{B}{1120 \pi^{3/2} R^3}  \Bigg[-35 A^3 \pi R^3 + 7 A^2 \sqrt{\pi} (40 + 3 C \sqrt{5 \pi}) R^3 
			- A (75 C^2 \pi + 280 \pi^2 + 112 C \sqrt{5 \pi}) R^3 
			+ 112 A (-3 B^2 + 5 \pi) (-1 + R)^3 \mathcal{F} \\
			&+ C \left(200 C \sqrt{\pi} R^3 + 5 \sqrt{5} C^2 \pi R^3 \right. 
			\left. + 8 \sqrt{5} (7 \pi^2 R^3 + 2 (6 B^2 - 7 \pi) (-1 + R)^3 \mathcal{F})\right)\Bigg],
		\end{split}
		\\
		\begin{split}
			\frac{\md C}{\md t} =  &-\frac{1}{616} \sqrt{\frac{\pi}{5}} \Bigg[22 C (7 \sqrt{5} A + 5 C) + 616 C \sqrt{\frac{5}{\pi}} 
			- \frac{66 C (7 \sqrt{5} A^2 + 10 A C + 5 \sqrt{5} C^2)}{\pi^{3/2}} 
			+ \frac{C (77 \sqrt{5} A^3 + 165 A^2 C + 165 \sqrt{5} A C^2 + 125 C^3)}{\pi} \\
			&\quad - \frac{22 B^2 (7 A^2 - 10 \sqrt{5} A C + 5 C^2) (-1 + R)^3 \mathcal{F}}{\pi^2 R^3} \Bigg].
		\end{split}
	\end{align}
\end{scriptsize}

\subsection{4 mode approximation}
\begin{scriptsize}
	\begin{align}
		\begin{split}
			\frac{\md A}{\md t} = 	&-\frac{1}{1120 \pi^{3/2} R^4} \Bigg[-280 A^3 \sqrt{\pi} R^4 + 35 A^4 \pi R^4 
			+ 5 \left(15 D^4 \pi + 28 D^2 \pi^2 - 224 \pi^3 - 16 D^3 \sqrt{5 \pi}\right) R^4 
			+ 200 D^2 (-1 + R)^3 (3 C^2 (-1 + R) + B^2 R) \mathcal{F} \\
			&+ 4 A \left(-210 D^2 \sqrt{\pi} R^4 + 10 \sqrt{5} D^3 \pi R^4 + 280 \pi^{3/2} R^4 
			+ \sqrt{5} D (-1 + R)^3 \left(15 C^2 (-1 + R) - 28 B^2 R\right) \mathcal{F}\right) \\
			&+ 70 A^2 \left(\pi \left(3 D^2 + 2 \pi\right) R^4 + 2 (-1 + R)^3 \left(3 C^2 (-1 + R) + 2 B^2 R\right) \mathcal{F}\right)\Bigg],
		\end{split}
		\\
		\begin{split}
			\frac{\md B}{\md t}  = &-\frac{1}{1120 \pi^{3/2} R^4} B \Bigg[35 A^3 \pi R^4 - 7 A^2 \sqrt{\pi} (40 + 3 D \sqrt{5 \pi}) R^4 
			+ A \left(75 D^2 \pi + 280 \pi^2 + 112 D \sqrt{5 \pi}\right) R^4 
			+ A (-1 + R)^3 \left(112 (3 B^2 - 5 \pi) R + 15 C^2 (-45 + 64 R)\right) \mathcal{F} \\
			&+ D \left(-200 D \sqrt{\pi} R^4 - 5 \sqrt{5} D^2 \pi R^4 
			+ \sqrt{5} \left(-56 \pi^2 R^4 + (-1 + R)^3 \left(C^2 (195 - 300 R) + 16 (-6 B^2 + 7 \pi) R\right) \mathcal{F}\right)\right)\Bigg],
		\end{split}	
		\\
		\begin{split}
			\frac{\md C}{\md t} = &-\frac{1}{4928 \pi^{3/2} R^4} C \Bigg[154 A^3 \pi R^4 + 11 A^2 \sqrt{\pi} (-112 + 3 D \sqrt{5 \pi}) R^4 
			+ 44 A \left((15 D^2 \pi + 28 \pi^2 - 4 D \sqrt{5 \pi}) R^4 + (-1 + R)^3 \left(4 (15 C^2 - 14 \pi) (-1 + R) \right.\right.\\
			&\left.\left. + B^2 (-19 + 64 R)\right) \mathcal{F}\right) 
			+ D \left(-1760 D \sqrt{\pi} R^4 + 95 \sqrt{5} D^2 \pi R^4 
			+ 4 \sqrt{5} \left(22 \pi^2 R^4 - (-1 + R)^3 \left(4 (-20 C^2 + 11 \pi) (-1 + R) + 11 B^2 (-7 + 20 R)\right) \mathcal{F}\right)\right)\Bigg],
		\end{split}
		\\
		\begin{split}
			\frac{\md D}{\md t} = &-\frac{1}{2464 \sqrt{5} \pi^{3/2} R^4} \Bigg[308 \sqrt{5} A^3 D \pi R^4 - 22 A^2 \left(6 D (-5 D \pi + 14 \sqrt{5 \pi}) R^4  
			- 15 C^2 (-1 + R)^4 \mathcal{F} + 28 B^2 (-1 + R)^3 R \mathcal{F}\right) \\
			&+ 2 D \left(2 \sqrt{\pi} (-330 \sqrt{5} D^2 + 125 D^3 \sqrt{\pi} + 616 \sqrt{5} \pi + 110 D \pi^{3/2}) R^4  
			+ 5 D (-1 + R)^3 \left(285 C^2 (-1 + R) - 44 B^2 R\right) \mathcal{F}\right) \\
			&+ 44 A D \left(-60 D \sqrt{\pi} R^4 + 15 \sqrt{5} D^2 \pi R^4  
			+ 2 \sqrt{5} \left(7 \pi^2 R^4 + 10 (-1 + R)^3 \left(3 C^2 (-1 + R) + B^2 R\right) \mathcal{F}\right)\right)\Bigg].
		\end{split}
	\end{align}
\end{scriptsize}

\section{Numerical evaluation of the velocity gradient}
Here we provide details of our numerical method. We emphasize that for the evaluation of the traction jump on $\partial S$ and for the evolution of the polarity field on $\partial D$ we only need to compute $\nabla \bu\big|_{\partial D}$. However, the traction jump $\bar{\bff}(\by)$ itself depends on $\nabla \bu\big|_{\partial D}$.  Treating this dependence explicitly makes the computations challenging for high MT bed density $\bar{\rho}$. As a result, here we adopt an implicit scheme as outlined below. Recall the integral representation for the velocity field is given as
\begin{equation}\label{eq:intfor}
	\bu(\bx) = \mathcal{S}[\bar{\bff}]\Big|_{\partial S \to \bx} -  \mathcal{S}[\bar{\bff}^w]\Big|_{\partial D \to \bx}.
\end{equation}
The wall-traction $\bff^w$ is evaluated using the no-slip boundary condition as follows:
\begin{equation}
	\bar{\bff}^{w,3 N \times 1} = - \left(\mathcal{S}^{-1}_{{\partial D} \to {\partial D}}\right)^{3 N \times 3N} \left(\mathcal{S}_{{\partial S} \to {\partial D}}\right)^{3 N \times 3N} \cdot \bar{\bff}^{3 N \times 1} = \mathcal{B}^{3 N \times 3 N} \cdot \bar{\bff}^{3 N \times 1}.
\end{equation}
where the spherical surface is discretized in $N$ points. The velocity gradient on the boundary is evaluated using a finite difference approximation. This can be represented as
\begin{equation}\label{eq:grad2}
	\left(\nabla \bu \right)^{9 N \times 1}\big|_{\partial D} = \mathcal{D}^{9N \times 3N} \cdot \left[\left(\mathcal{S}_{\partial S \to \partial D}\right)^{3 N \times 3 N}  + \mathcal{B}^{3 N \times 3 N} \right] \cdot \bar{\bff}^{3N\times 1} = \mathcal{C}^{9N \times 3N} \cdot \bar{\bff}^{3N\times 1},
\end{equation}
where $\mathcal{D}$ is the finite difference matrix. We now write our traction jump as follows
\begin{equation}
	\bar{\bff}(\by) = \underbracket{\bar{\rho} \bar{\sigma} \bn + \bar{\rho} \chi \bT_0 \times \bn}_{\bff_a} - \underbracket{{\bar{\rho} \chi} \frac{\mathbf{n n n}}{2}: \nabla \mathbf{u}}_{\bff_c} .
\end{equation}
We note that it is possible to write
\begin{equation}
	\bff_c= \mathcal{A}(\bn)^{3N \times 9N} \cdot 	\left(\nabla \bu \right)^{9 N \times 1}.
\end{equation}
Using this notation, we find the velocity gradient on the boundary is given as the solution of the following linear system
\begin{equation}
	\left[\mathbb{I} + \frac{\bar{\rho} \chi}{2} \mathcal{C}^{9 N \times 3N} \cdot \mathcal{A}(\bn)^{3N \times 9N} \right] \cdot \left(\nabla \bu \right)^{9 N \times 1}  = \mathcal{C}^{9N \times 3N} \cdot \bar{\bff}_a^{3N\times 1}.
\end{equation}
This is solved using GMRES which typically takes 6-7 iterations to converge. The polarit field is then evolved using a 2nd order Runge Kutta method. We further highlight that since the geometry is fixed, the discrete operator $\mathcal{C}$ can be precomputed at the start of the simulation. 

\bibliography{bibSI}